Research Article

**Title:** The shrinking human protein coding complement: are there now fewer than 20,000 genes?

**Authors:** Iakes Ezkurdia[1*], David Juan[2*], Jose Manuel Rodriguez[3], Adam Frankish[4], Mark Diekhans[5], Jennifer Harrow[4], Jesus Vazquez[6], Alfonso Valencia[2,3], Michael L. Tress[2,*].

**Affiliations:**
1. Unidad de Proteómica, Centro Nacional de Investigaciones Cardiovasculares, CNIC, Melchor Fernández Almagro, 3, rid, 28029, MadSpain
2. Structural Biology and Bioinformatics Programme, Spanish National Cancer Research Centre (CNIO), Melchor Fernández Almagro, 3, 28029, Madrid, Spain
3. National Bioinformatics Institute (INB), Spanish National Cancer Research Centre (CNIO), Melchor Fernández Almagro, 3, 28029, Madrid, Spain
4. Wellcome Trust Sanger Institute, Wellcome Trust Campus, Hinxton, Cambridge CB10 1SA, UK
5. Center for Biomolecular Science and Engineering, School of Engineering, University of California Santa Cruz (UCSC), 1156 High Street, Santa Cruz, CA 95064, USA
6. Laboratorio de Proteómica Cardiovascular, Centro Nacional de Investigaciones Cardiovasculares, CNIC, Melchor Fernández Almagro, 3, 28029, Madrid, Spain

*: these two authors wish to be considered as joint first authors of the paper.

**Corresponding author:**
Michael Tress, mtress@cnio.es,
Tel: +34 91 732 80 00
Fax: +34 91 224 69 76

**Abstract**

Determining the full complement of protein-coding genes is a key goal of genome annotation. The most powerful approach for confirming protein coding potential is the detection of cellular protein expression through peptide mass spectrometry experiments. Here we map the peptides detected in 7 large-scale proteomics studies to almost 60% of the protein coding genes in the GENCODE annotation the human genome. We find that conservation across vertebrate species and the age of the gene family are key indicators of whether a peptide will be detected in proteomics experiments. We find peptides for most highly conserved genes and for practically all genes that evolved before bilateria. At the same time there is almost no evidence of protein expression for genes that have appeared since primates, or for genes that do not have any protein-like features or cross-species conservation. We identify 19 non-protein-like features such as weak conservation, no protein features or ambiguous annotations in major databases that are indicators of low peptide detection rates. We use these features to describe a set of 2,001 genes that are potentially non-coding, and show that many of these genes behave more like non-coding genes than protein-coding genes. We detect peptides for just 3% of these genes. We suggest that many of these 2,001 genes do not code for proteins under normal circumstances and that they should not be included in the human protein coding gene catalogue. These potential non-coding genes will be revised as part of the ongoing human genome annotation effort.

**Introduction**

The actual number of protein coding genes that make up the human genome has long been a source of discussion. Before the first draft of the human genome came out many researchers believed that the final number of human protein coding genes would fall somewhere between 40,000 and 100,000 (Pennisi 2003). The initial sequencing of the human genome revised that figure drastically downwards by suggesting that the final number would fall somewhere between 26,000 (Venter 2001) and 30,000 (International Human Genome Sequencing Consortium 2001) genes. With the publication of the final draft of the Human Genome Project (International Human Genome Sequencing Consortium 2004) the number of protein coding genes was revised downwards again to between 20,000 and 25,000. Most recently, Clamp and co-workers (Clamp *et al* 2007) used evolutionary comparisons to suggest that the most likely figure for the protein coding genes would be at the lower end of this continuum, just 20,500 genes.

The Clamp analysis suggested that a large number of ORFs were not protein coding because they had features resembling non-coding RNA and lacked evolutionary conservation. The study suggested that there were relatively few novel mammalian protein-coding genes and that the approximately 24,500 genes annotated in the human gene catalogue would end up being cut by 4,000.

The Ensembl project began the annotation of the human genome in 1999 (Hubbard *et al* 2002). The number of genes annotated in the Ensembl database (Flicek *et al* 2013) has been on a downward trend since its inception. Initially there were more than 24,000 human protein-coding genes predicted for the reference genome, but that number has gradually been revised lower. More than two thousand automatically predicted genes have been removed from the reference genome as a result of the merge with the manual annotation produced by the Havana group (Harrow *et al* 2006), often by being re-annotated as non-coding biotypes. The numbers of genes in the updates of merged GENCODE geneset is now close to the number of genes predicted by Clamp in 2007. The most recent GENCODE release (GENCODE 18) contains approximately 20,700 protein-coding genes.

The GENCODE consortium is composed of nine groups that are dedicated to producing high accuracy annotations of evidence-based gene features based on manual curation, computational analyses and targeted experiments. The consortium initially focused on 1% of the human genome in the Encyclopedia of DNA Elements (ENCODE Project Consortium 2012) pilot project (Harrow *et al* 2006**,** ENCODE Project Consortium 2007) and expanded this to cover the whole genome (Harrow *et al* 2012).

Manual annotation of protein-coding genes requires many different sources of evidence (Guigó *et al* 2006, Harrow *et al* 2012). The most convincing evidence, experimental verification of cellular protein expression, is technically challenging to produce. Although some evidence for the expression of proteins is available through antibody tagging (Uhlen *et al* 2010) and individual experiments, high-throughput tandem mass spectrometry (MS)-based proteomics methods are the main source of evidence. Proteomics technology has improved considerably over the last two decades (Aebersold and Mann 2003, Mallick and Kuster 2010) and these advances are making MS an increasingly important tool in genome annotation projects. High quality proteomics data can confirm the coding potential of genes and alternative transcripts, and this is especially useful in those cases where there is little additional supporting evidence. A number of groups have demonstrated how proteomics data might be used to validate translation (Tanner *et al* 2007, Brosch *et al* 2011, Ezkurdia *et al* 2012). However, while MS evidence can be used to verify protein-coding potential, the low coverage of proteomics experiments implies that the reverse is not true. Not detecting peptides does not prove that the corresponding gene is non-coding since it may be a consequence of the protein being expressed in few tissues, having very low abundance, or being degraded quickly.

Finding peptides for all protein coding genes is the holy grail of proteomics, and a number of recent large-scale experiments have detected protein expression for approximately 50% of the human genome (Beck *et al* 2011, Muñoz *et al* 2011, Nagaraj *et al* 2011, Ezkurdia *et al* 2012, Geiger *et al* 2012, Kristensen *et al* 2013, Neuhauser *et al* 2013). However, this is still some way short of the 97% claimed for the single-celled *Saccharomyces cerevisiae* (Picotti *et al* 2013).

Here we put together reliable peptide evidence from seven separate large-scale MS analyses and confirm protein expression for 11,838 protein-coding genes. We show that the vast majority of these confirmed protein-coding genes correspond to the oldest and most conserved ORFs. We also describe a set of 2,001 genes that have little or no peptide evidence from the seven proteomics studies and that have multiple features that suggest that they may not code for proteins. These results lead us to conclude that the human genome is likely to have fewer than 20,000 protein-coding genes.

**Results**

We collected peptides from seven separate MS sources. Two came from large-scale proteomics databases, PeptideAtlas (Farrah *et al* 2013) and NIST (http://peptide.nist.gov/). Another four, referred to as "Geiger", "Muñoz", "Nagaraj" and "Neuhauser" throughout the paper, were recently published large-scale MS experiments (Muñoz *et al* 2011, Nagaraj *et al* 2011, Geiger *et al* 2012, Neuhauser *et al* 2013). For all six datasets the starting point was the list of peptides provided by the authors or databases. We generated the final set of peptides (referred to as "CNIO") in house from an X!Tandem (Craig and Beavis 2003) search against spectra from the GPM (Craig *et al* 2004) and PeptideAtlas databases, following the protocol set out in Ezkurdia *et al* (Ezkurdia *et al* 2012) with a false discovery rate of 0.1%. These seven studies cover a wide range of tissues and cell types.

In order to improve reliability the peptides from each of these studies were filtered, eliminating non-tryptic and semi-tryptic peptides and peptides containing missed cleavages. For those studies where it was possible we considered only peptides identified by multiple search engines. We identified a total of 255,188 peptides in the seven analyses and mapped them to genes in the GENCODE 12 annotation of the human genome (GENCODE 12 corresponds to Ensembl 67). Peptides were separated into discriminating peptides, those that mapped uniquely to a single gene in the annotation, and non-discriminating peptides, those that matched more than one gene product (a total of just over 14,000 peptides).

We considered a gene uniquely identified when at least two discriminating peptides matched the gene. Using these criteria we identified 11,840 genes, **57.9%** of the genes annotated in GENCODE 12. We were able to map non-discriminating peptides to another 1,648 genes, while 6,974 genes had no peptide evidence at all.

The number of genes detected is comparable with previous recent studies (Beck *et al* 2011, Munoz *et al* 2011, Nagaraj *et al* 2011, Ezkurdia *et al* 2012, Geiger *et al* 2012, Kristensen *et al* 2013, Neuhauser *et al* 2013). There was a substantial overlap between the seven datasets – peptides for 5,033 genes were detected in all datasets and there were indications that this number would have been higher since the Nagaraj (Nagaraj *et al* 2011), Geiger (Geiger *et al* 2012) and Neuhauser (Neuhauser *et al* 2013) experiments excluded a number of highly expressed proteins (such as Titin and Albumin). 9,781 genes were identified in four or more different datasets.

The PeptideAtlas collection identified the highest number of genes (10,394 genes from 127,404 peptides). Surprisingly the combination of the seven datasets did not substantially increase the number of genes detected – the addition of the six other datasets to the PeptideAtlas peptides identified just 1,444 additional genes. However, the identification of peptides across multiple experiments did serve to increase the confidence of the identifications. Our results suggest that the detected and undetected genes in proteomics experiments form two relatively robust and well-defined groups.

**The relationship between proteomics detection and gene features**

We were interested in determining the reasons for not detecting peptides for the other 40% of human genes. Three technical reasons make protein detection more difficult *a priori*. Firstly the length of the protein influences the probability of peptide identification since the shorter the protein is, the fewer peptides can be produced making identification of low molecular weight proteins technically challenging. The range of transcript expression is also important; if a transcript is expressed in very few tissues, it is less likely to be detected at the level of protein. Finally, it is well known that it is more difficult to detect peptides for proteins with trans-membrane helices because membrane-bound proteins are poorly accessible to tryptic digestion and hydrophobic peptides may be difficult to detect in conventional reverse-phase columns. Therefore we first analysed these three factors in detail.

We found a relationship between protein length and peptide detection (**supplementary fig 1**). Very few peptides were detected for proteins shorter than 50 amino acid residues; in fact we did not detect peptides for proteins shorter than 38 residues.

We reasoned that it should be easier to detect peptides for genes that express transcripts across many tissues. Using data from UniGene (Wheeler *et al* 2003) we found a strong correlation between peptide detection and the number of tissues in which a transcript was expressed (**Supplementary Fig 2**). We detected protein evidence for over 90% of the 6,286 genes that express transcripts in 24 or more tissues. In contrast we detected peptides for fewer than 25% of the 2,932 genes that express transcripts in two or fewer tissues.

In order to measure the effect of trans-membrane helices on peptide detection we culled the trans-membrane helix predictions from the APPRIS database (Rodriguez *et al* 2013). We compared detection rates for proteins that contained trans-membrane helices against those that did not. The results (**Supplementary Figure 3**) confirmed that proteomics detects fewer peptides for proteins with *multiple* trans-membrane helices (we found peptides for just 39.1% of these genes). Proteins with single trans-membrane helices were just as likely to be detected as those without trans-membrane helices.

Genes with low molecular weight gene products, with restricted transcript expression, or with multiple trans-membrane helices made up over 3,400 of the 7,000 genes that we did not identify, so these three features go some way towards explaining why we only detect peptides for just 58% of the protein coding genes. The olfactory receptors are a good example. These proteins have multiple trans-membrane helices and their expression is tissue-restricted (though, curiously, UniGene registers expression in as many as 17 different tissues). We do not detect any peptide evidence for *any* of the 380 diverse olfactory receptors annotated in GENCODE 12.

Following the trans-membrane helix comparison we investigated the effect of a range of protein features on protein detection rates. Again we collected features from the APPRIS database and measured peptide detection rates against the presence and absence of features such as protein functional domains, functional residues, homology to known structures and cross-species conservation. Proteins with these features were more likely to be detected in the analyses than proteins without these features (**supplementary figure 3**). For example, we detected 75.3% of proteins annotated with catalytic or ligand binding residues and 73.9% of proteins annotated with at least one *unbroken* PfamA functional domain.

In addition, we discovered that genes without any APPRIS protein features had very low rates of peptide detection. In fact the absence of all protein-like features turned out to have a very strong inverse relationship to peptide detection - we detected peptides for just 4.2% of the 956 genes that did not have protein-like features or conservation in APPRIS. By way of contrast we detected peptide evidence for almost 1,500 (30%) of those genes with low molecular weight products, restricted expression or multiple trans-membrane helices.

**Conservation and gene age are the best predictors of peptide detection**
While the absence of protein-like features was the best predictor of *non-detection* in proteomics experiments, we found that the best predictors of whether a protein will be detected in proteomics experiments were coding sequence conservation and the first ancestral species in which an ancestral gene is detected for the gene family (gene family age).

To look at the effect of conservation on gene detection, we collected data from INERTIA (Rodriguez *et al* 2013), one of the modules of the APPRIS database. INERTIA generates scores for the evolutionary rates of codons and exons for splice variants. Evolutionary rates in INERTIA are calculated using SLR (Massingham and Goldman 2005) and multiple alignments of orthologous vertebrate transcripts (Lindblad-Toh *et al* 2011). We defined gene conservation from the INERTIA score of the most conserved exon (MI score).

We found a striking correlation between conservation and detection in proteomics experiments (**Figure 1**). We detected peptides for 84.7% of the 5,554 genes with an MI score below 0.02 (the most conserved genes); in contrast we detected little evidence for the 992 genes with least protein coding-like conservation (MI score > 0.8, 6.1% detection). There were 575 genes that had tiny exons or alignments against fewer than three species, so had no MI score. For these genes the detection rate was less than 2%. Genes with poor conservation for which we could not find protein structural or functional features were hardly detected at all (**Figure 1**).

Gene family age (the oldest phylogenetic division that has a gene from the same family) was calculated using phylogenetic trees from Ensembl Compara (Flicek *et al* 2011). Although there is a relation between gene family age and conservation, they are not exactly the same. INERTIA conservation is calculated only from alignments of vertebrates, while gene family age is measured from the Fungi-Metazoan period. A gene may have an older gene family age, for example Bilateria, but the gene itself may have arisen from a primate duplication. Thus there are genes with relatively recent gene family age and well-conserved exons, and many genes with the earliest gene family ages (Fungi-Metazoa, Bilateria, Coelomata) and poor MI scores.

We compared proteomics detection rates for each of the gene family ages. The results are in **figure 2**. We detect peptides for over 89% of the genes from the oldest phylogenetic division (those that have Fungi-Metazoa group family age), while we detect practically no peptides for those genes whose family age can only be traced back to primates.

We also determined gene age, the phylogenetic division in which the most recent ancestral duplication occurred. The results for gene age show a similar trend to family age (results are shown in **supplementary figure 4**).

Combining gene family age with gene age or conservation gives even more striking results. We detect peptides for 96.4% of genes with both Fungi-Metazoa family age and Fungi-Metazoa gene age (1,136 genes), and 96.5% of highly conserved Fungi-Metazoa family age genes (1,712 Fungi-Metazoa genes with MI scores below 0.015).

In order to determine whether the link between gene family age and proteomics detection was confined to humans, we performed a similar experiment using yeast, a single-celled organism. We generated gene ancestral definitions by all against all sequence similarity searches and plotted gene family age against the percentage of genes for which peptides are recorded in PeptideAtlas. There is the same clear relationship between gene age and detection rates in yeast (**supplementary figure 5**) so there were no peptides for those genes belonging to families first detected in *cerevisiae*.

**A set of potential non-coding genes**
The results from the conservation and gene family age analyses show that the most recently evolved genes (those with primate gene family age), the least conserved genes and genes without protein features were much less likely to be detected in proteomics experiments (just 0.9% of the 563 primate-specific genes, and 2% of the 987 genes with MI scores greater than 1 were detected with discriminating peptides).

We searched a range of sources, including the APPRIS database, the UniProt protein database (UniProt Consortium 2013) and Ensembl/GENCODE, to find other features that might be related to low detection rates. From these we selected a list of 19 features that correlated with low protein detection rates (**see table 1,** and the methods and **supplementary** sections for more details).

We produced a set of 2,001 genes that had at least one of these 19 features. We detected peptides for just 61 (3%) of these genes. The combination of features not typical of proteins and the very low peptide detection rates suggested that a number of these genes might turn out to be non-coding genes or pseudogenes.

Many genes in the set had more than one of the features listed in **table 1**, and the more features a gene had, the less likely it was to be identified in the proteomics analysis. We found no peptides at all for genes with five or more features (**supplementary fig 6**).

Almost half the genes in this potential non-coding set were annotated with clone-based names rather than function-based names typical of protein coding genes. Many of those with non-clone names were named for their chromosomal position, their sequence bias or with one of a set of miscellaneous identifiers that included generic names, pseudogene names, chimeric gene names, and the cutely named "orphan" gene (**supplementary fig 7**).

Immunoglobulin and t-cell segments, keratin-binding proteins, various antigens and olfactory receptors made up almost 300 genes. Analysis with the DAVID functional annotation tool (Huang da *et al* 2009) tool bears this out. Those genes we could map had biases towards the GO terms "intermediate filament" (Benjamini score 5.3e-20), "keratinization" (4.1e-10), "defence response to bacterium" (5.2e-14) and "extra-cellular region" (5.9e-6, a number of genes in the set are secreted). DAVID only identified 50% of genes in the potential non-coding set.

The set had two other biases. First, there were a number of genes with human-specific duplications. Clearly it was impossible to distinguish these genes because most do not have unique tryptic peptides. Multiple duplications often generate pseudogenes, so some duplicated genes will be non-functional. Second, there were 142 genes corresponding to proteins that were 38 residues or shorter. Many of these were immunoglobulin or T-cell receptor joining segments and some were annotated as pseudogenes.

**Do these genes code for proteins?**

In order to determine whether the genes in the potential non-coding set coded for proteins, we split the GENCODE 12 genes into 3 groups, those genes for which we found peptides in the 7 analyses (*Detected*), the 1,940 genes in the potential non-coding/pseudogene set for which we did not detect peptides (*Potential NC*) and the remaining genes, those genes for which we did not find peptides, but that were not in the potential non-coding set (*Not Detected*). We looked at the transcript evidence from UniGene for the three sets. We counted the number of tissues in which each gene had evidence of 5 or more transcripts per million. The results show that there is much more transcript evidence for the *Detected* than for the *Potential NC* genes (**figure 3**). Although distributions of transcript expression for the *Not Detected* and *Potential NC* sets are similar, it is noticeable that more than 50% of the genes in the *Potential NC* set have no measured transcription in any of the 45 tissues in UniGene.

In 2007 Clamp (Clamp *et al* 2007) identified 1,177 genes as "orphans" with features typical of non-coding RNA. Based on the numbers of coding genes then available (Ensembl 35), the authors suggested that the human reference genome had only 20,500 coding genes. The GENCODE 12 annotation still annotates 308 of these orphans and there were 248 orphans among the potential non-coding genes. We detected peptides for just 3 of these genes.

As part of their analysis the authors calculated a reading frame conservation (RFC) score for transcripts. Reading frames that change in pairwise alignments between two different species suggest large changes in protein function. The RFC score distribution of the orphans was very close to that of non-coding RNA.

We carried out our own reading frame analysis. RFC scores were generated for the known protein coding genes in the *Detected* set, for the probable protein coding genes in the *Undetected* set, for the genes in the *Potential NC* set and for a set of non-coding RNA. We aligned human with four species, chimp, macaque, dog and mouse, as in the Clamp analysis.

The results are shown in **figure 4**. For the human-chimp alignments most transcripts keep the same frame over the whole alignment – this is true even for the set of the non-coding genes we analysed – so the differences between the *Detected* genes and the genes in the *Potential NC* set were minimal. With the human-macaque alignments the frame is lost in most non-coding alignments, and there is a marked difference between the *Detected* and *Potential NC* genes, while in the case of dog and mouse practically all non-coding transcript alignments have frame changes and the proportion of the *Potential NC* transcripts that change frame is approximately half that of the *Detected* transcripts.

The RFC results for mouse and dog suggest that between half and three quarters of the genes in the *Potential NC* set are unlikely to code for proteins. We could not find dog or mouse orthologues for half of the genes in the *Potential NC* set. This is the same proportion as the non-coding genes, which suggests that as many as half the genes in the *Potential NC* set are non-coding or orphan protein coding genes. For those transcripts we *were* able to align (**figure 4**) half had changes in reading frame. While this is not as bad as the non-coding set (practically all non-coding genes had frame changes in the human-dog and human-mouse alignments), it does suggest that another 25% of the genes in the *Potential NC* set will have dramatic changes in function compared to their mouse and dog orthologues, which again is not indicative of protein coding potential.

The RFC results are also consistent with a model in which the genes in the *Not Detected* set are protein coding. Their RFC scores are close to those of the known protein coding genes in all four sets of alignments. The fact that many of these genes are expressed in limited tissues (see **figure 3**) is likely to be part of the reason these genes were not detected in proteomics analyses.

We generated four sub-groups (read-throughs, possible coding, possible non-coding, possible pseudogenes) from the potential non-coding set by manual curation. We repeated the RFC analysis on these 4 groups (**supplementary figure 8**). Possible protein coding genes and read-through genes had RFC scores that were very similar to known protein coding genes, while the possible non-coding genes had similar RFC scores to known non-coding genes and the possible pseudogenes had RFC scores that were somewhere between the two.

**Can we detect expression of proteins by other means?**
In order to explore whether further tissue specific proteomics experiments are likely to modify our findings we looked for evidence of the genes in the potential non-coding set in other sources, first in a large-scale proteomics experiment using human placental tissue (Lee *et al* 2013). The authors detected peptides that mapped to 4,239 genes, including 51 that were not identified in our analysis. However, all 51 genes were in the *Not Detected* set. Fifteen of these new genes had evidence for expression in placenta, suggesting that large-scale proteomics experiments performed on specific tissues may detect gene products with restricted expression.

Since large-scale MS analyses are not the only form of confirming protein expression, we attempted to detect protein evidence by other means for the 2,001 genes in the potential non-coding set. The Passel initiative (Farrah *et al* 2012) is a data repository obtained from targeted proteomics experiments performed using selected reaction monitoring (SRM) measurements. The SRM approach is used in many proteomics experiments in order to improve the detection of low-abundance proteins.

We interrogated the Passel resource and found evidence for five genes in the potential non-coding set, including two (*TSPO* and *SPRR3*) for which we already had detected peptides.

The Human Protein Atlas (Uhlen *et al* 2010) is a resource that makes use of antibody-based proteomics to catalogue the components of the human proteome. We were only able to identify unique 21 genes (and one pair of genes, *DEFA1* and *DEFA1B*, that were protein sequence identical) with a good or medium reliability scores. Of these, 13 had a high Human Peptide Atlas reliability score, and a further 10 had a medium reliability score. We had detected peptides for three of these 22 genes in the seven studies.

The UniProt literature resource (UniProt Consortium 2013) for each gene provides links to experimental papers for each gene wherever possible. We investigated the links for the genes in this set and found that there is evidence of the function or expression of the protein for 46 separate genes. We had found peptides for 13 of these genes and there was evidence for two other genes in the Human Peptide Atlas resource.

In total using Passel, Human Protein Atlas and UniProt we were only able to turn up evidence of protein expression for a further 54 (2.7%) of the 2,001 genes in the potential non-coding set (see **Fig 5**). We were not able to find any evidence of protein expression for 1,886 genes in the potential non-coding set.

**Discussion**

Our analysis of seven large-scale proteomics experiments has unambiguously identified close to 12,000 human genes. We found most peptides for the oldest and most conserved genes. The high proportion of ancient genes identified with peptide evidence is in accord with their expected expression level and importance to the cell. Ancient genes are generally widely expressed and often retain important housekeeping roles. We identified 96.3% of genes that have not duplicated since the Fungi-Metazoan era.

The absence of peptides in proteomics experiments does not necessarily imply that a protein is not expressed, but the high coverage of these 7 studies lead us to analyse the 40% of genes that were not identified in the analysis. The 7 proteomics studies covered wide range of cell types, so one of the main reasons for not detecting a protein, that it is expressed in limited tissues or developmental stages, should not be so important. The PeptideAtlas database alone is a compendium of experiments carried out on 51 different tissue and cell types, and the PeptideAtlas database forms just a part of the CNIO study and the NIST database. Six of the seven studies were carried out on a range of tissues and together these studies cover considerably more cell types than UniGene.

Despite the range of tissues interrogated, it is still probable that some proteins not detected because they are tissue specific. In addition there may also be proteins that are expressed in very low quantities or, like the *HOX* genes, have a very short half-lives. Some proteins also have features, such as multiple-trans-membrane helices, that make them difficult to detect for technical reasons.

There is one other potential explanation for not identifying peptides for a gene. It may be that some genes do not actually code for proteins. To investigate this we selected 2,001 genes that had one or more features that suggested of a lack of coding potential. We found protein evidence for less than 6% of these genes and for many of these genes the reading frame was not conserved in cross-species alignments. Together the non-protein like features, loss of reading frame and lack of protein evidence suggested that many of these genes might not code for proteins under normal circumstances.

We did detect proteomics evidence for several genes in this set, for example *SLC5A3*, which is annotated by UniProt as only having evidence of protein existence by homology, and for *SPA17*, which is annotated as "putative" by GENCODE. Both these genes have protein-like features and good conservation. They were in the list of potential non-coding genes because the human genome annotation project is not yet complete and these genes had yet to be annotated with evidence by GENCODE and UniProt. *SPA17* is no longer tagged as "putative" in the GENCODE 18 annotation. We also identified peptides for genes with conflicting protein coding evidence, such as *WASH4P* and

*WASH6P*, annotated as protein coding, but tagged as pseudogenes in the Ensembl description. Again as the human genome annotation progresses, these descriptions are likely to be refined.

There are genes that do not fit into the conventional coding/non-coding narrative in the non-coding set. Several genes are annotated as potentially non-functional, but may actually be functional under certain conditions. One example is *FMO2*, Dimethylaniline monooxygenase 2. There are two alleles, FMO2*2A, which is truncated, and FMO2*1, the full-length form. The truncated allele FMO2*2A is catalytically inactive and is probably unable to fold correctly. FMO2*1 is not present in Caucasian and Asian populations, but is found in low quantities in African populations (Veeramah *et al* 2008). The function of the FMO2*1 variant is not clear, but it does lead to increased risk of thiourea-*caused* pulmonary toxicity.

The growth of the number of annotated read-through genes plays a role in maintaining artificially high numbers of protein coding genes in the human reference genome. Read-through genes connect two or more neighbouring genes by splicing together exons of two otherwise defined, independent loci. It is not clear what biological significance this has, but the number of read-through genes is growing in the reference annotation. We found 229 read-through genes in GENCODE 12 and there are 407 read-through protein-coding genes annotated in GENCODE 18. There is some evidence to suggest that read-through transcription is part of a process that allows genes to gain new protein domains (Buljan *et al* 2010) and so these genes might be regarded testing ground for new protein functions. However, there is very little peptide evidence for these chimeric genes (just 0.87%) and they are probably best annotated as splice variants of the downstream gene. Their presence in the reference genome makes proteomics searches (and other large-scale experiments) more complicated because it is impossible to find peptides that separate the (likely non-coding) read-through genes from the component genes. Without the read-through genes we could have identified up to 300 more genes in the 7 proteomics studies.

Many of the 2,001 genes in the potential non-coding set may turn out not to code for proteins under any circumstances. Unfortunately, genes annotated as protein coding at the gene annotation level can have complications for downstream services and research groups that are sometimes difficult to undo. The Pfam functional domain database, for example, has a recent surge in the numbers of newly defined protein functional domains and many of these have almost certainly been defined on the back of "protein-coding" genes, some of which may turn out not to code for proteins. Over-estimating the numbers of protein coding genes can also hinder experiments such as large-scale proteomics projects and biomedical projects, such as the mapping of cancer or disease-related variations to human genes.

The human genome is still in the process of being annotated and the Ensembl/GENCODE merge of the human genome is in constant flux as the annotators withdraw, redefine gene models and add new genes. To some extent our results reflect this situation, many of the genes we have identified will be removed from the protein-coding catalogue as the manual annotations become more complete. In fact this can be seen clearly with the most recent release of the reference annotation, GENCODE 18, where 328 of the 2,001 genes in the potential non-coding set have already been withdrawn or redefined.

Most genes in the potential non-coding set have multiple non-coding features, little or no evidence of transcript expression, no detected peptides, and a reading frame conservation that fits non-coding genes more closely than coding genes. We believe that this evidence suggests that as many as 1,500 genes do not code for proteins. Our evidence suggests that the final number of true protein coding genes in the reference genome may lie closer to 19,000 than to 20,000.

**Methods**

Peptides were assembled from seven previously available proteomics datasets. Four of the peptide datasets, the Geiger, Muñoz, Nagaraj and Neuhauser sets, came from published large-scale experiments (Muñoz *et al* 2011, Nagaraj *et al* 2011, Geiger *et al* 2012, Neuhauser *et al* 2013), two others were large spectra libraries, PeptideAtlas (Farrah *et al* 2013) and NIST (http://peptide.nist.gov/). The final study (referred to as CNIO throughout the paper) was carried out in-house and is detailed below.

**The CNIO analysis**

The CNIO analysis was based on the protocol detailed in Ezkurdia *et al*. (Ezkurdia *et al* 2012). Briefly we used X!Tandem (Craig and Beavis 2003) to search against peptide mass spectra from two publicly available proteomics resources, the Global Proteome Machine Organization (GPM, Craig *et al*. 2003) and PeptideAtlas databases. We used an updated set of spectra in the analysis; the spectra data in the GPM database had grown by 37% and PeptideAtlas by 18% since the original experiment. The PeptideAtlas and GPM data files can be downloaded from the Tranche distributed file system (tranche.proteomecommons.org) and ftp://ftp.thegpm.org/data/msms/.

Peptides were identified by searching against GENCODE 12. Expectation values (e-values) produced by X!Tandem were used to score the peptide-spectrum matches (PSM). When a peptide is identified more than once, we only included the PSM with lowest e-value. Only fully tryptic peptides containing a maximum of one missed tryptic cleavage site were taken into account. Peptides were considered positively identified when they had a FDR equal or below 0.1%. The FDR was calculated using a concatenated target/decoy strategy (Moore *et al* 2002); the decoy database was constructed by reversing each GENCODE 12 sequence entry.

**Filtering for high quality peptide identifications**

In order to guarantee that we only used the most reliable data from these sets, the peptides (and in the case of the CNIO study, the spectra) were filtered before mapping to the genome annotation. We used a series of filters to remove the most likely false positive peptides in each analysis.

It has been shown that using multiple search engines increases performance (Colaert *et al* 2011) so where possible we required peptides to be identified by more than one search engine. For the Nagaraj, Neuhauser and Geiger datasets we were able to use peptides with an Andromeda (Cox *et al* 2011) score of 100 or more, since this is the score above which Andromeda and Mascot (Koenig *et al* 2008) are almost always in agreement on the top-scoring peptides (Cox *et al* 2011).

The NIST database uses five different search engines to identify peptides from spectral databases. The NIST data has good coverage of the human genome but a higher than 1% FDR. We filtered the NIST peptides by only including peptide-spectra matches where three or more of the search engines identified the same peptide. The Muñoz study and PeptideAtlas database peptides did not have any specific filters.

An in-house investigation of the false positive rates of the various types of peptides showed that non-tryptic peptides, semi-tryptic peptides and missed cleavages without the presence of one the cleaved tryptic sub-peptides had markedly higher false positive rates. Non-tryptic peptides, semi-tryptic peptides and peptides with unsupported missed cleavages were removed from all sets. We applied the equivalent rule to peptides detected using GluC and LysC enzymes in the Nagaraj analysis (Nagaraj *et al* 2011).

We mapped the peptides to the GENCODE 12 geneset. The manual GENCODE annotations for Ensembl annotation are probably the most reliable annotation of human protein coding genes. The version we used was GENCODE 12 (equivalent to Ensembl 67), which is annotated with 20,462 protein-coding genes. We counted both the number of peptides that mapped unequivocally to a single gene (*discriminating peptides*) and those that mapped two or more different genes (*non-discriminating peptides*). In order to prove the expression of a protein we required two discriminating peptides or discriminating peptides from two or more analyses.

**Protein features**
Protein features were supplied by APPRIS (Rodriguez *et al* 2013), a database that houses annotations of protein structural and functional data and information from cross-species conservation for the human genome. Genes were annotated with protein structural information via a mapping to structural homologs in the PDB (Rose *et al* 2011), highly reliable predictions of conserved functionally important amino acid residues were made by *firestar* (Lopez *et al* 2011), mapping to Pfam functional domains was carried out via Pfamscan (Punta *et al* 2012). In addition trans-membrane helices were predicted using three separate trans-membrane predictors (Käll *et al* 2004, Viklund and Elofsson 2004, Jones 2007). We predicted signal peptides with SignalP (Emanuelsson *et al* 2007). Conservation information comes from two sources, one APPRIS module counts the numbers of equivalent vertebrate orthologues in the protein databases, while a second, INERTIA, calculates exon evolutionary rates using three separate sets of cross-vertebrate transcript alignments (Blanchette *et al* 2004, Lassmann *et a*. 2005, Löytynoja and Goldman 2005) and SLR (Massingham and Goldman 2005). APPRIS calculates features for all transcripts and we took the APPRIS scores from the highest scoring transcript for each gene.

**INERTIA MI scores (exon conservation scores)**

The results from INERTIA were used to calculate the conservation score for each gene, a score that is referred to as MI score throughout the paper. The MI score was the INERTIA score from the lowest scoring exon. There were two caveats: the exon had to be at least 42 bases long and the alignment had to have at least three species other than human.

**Gene Expression Breadth based on EST data**

Expression data were obtained from the UniGene database (Wheeler *et al* 2003) at http://www.ncbi.nlm.nih.gov/unigene/ (data download from August 2013). UniGene provides EST data clustered in different sets according to the different tissues (45 body sites). We considered that a gene is expressed if at least two cDNAs were found, representing 5 or more transcripts per million. Genes with no expression data within the tissue sets were removed from the corresponding analyses and as a consequence, we obtained 17,934 human genes with tissue expression information.

**Human Gene birth dating**

We performed a gene birth dating analysis based on phylogenetic family trees following a pipeline that is conceptually similar to that described recently (Roux and Robinson-Rechavi 2011). We used the phylogenetic reconstructions of ENSEMBL Compara v67 (Flicek *et al* 2011), which are based on genes sequenced from 58 different species. We focused on the human protein coding genes annotated by ENSEMBL Compara v67. We only considered age classes (or phylostrata) representing the last common ancestors of *Homo sapiens* and species sequenced with relatively high coverage (at least 5X). We decided to remove Euarchontoglires phylostratum and to collapse it within the Eutherian level due to the inconsistencies described previously between gene trees and species phylogeny at this level (Cannarozzi *et al* 2007, Huerta-Cepas *et al* 2007). Our analysis included the following 18 age classes for human genes: Fungi/Metazoa, Bilateria, Coelomata, Chordata, Vertebrata, Euteleostomi, Sarcopterygii, Tetrapoda, Amniota, Mammalia, Theria, Eutheria (Eutheria + Euarchontoglires), Simiiformes, Catarrhini, Hominoidea, Hominidae, HomoPanGorilla and Homo sapiens. For the purposes of the graphic in Figure 2 all classes from Simiiformes to Homo sapiens were combined to form the "Primate" class and the smaller classes Vertebrata and Sarcopterygii were clustered together with Chordata and Tetrapoda respectively.

ENSEMBL Compara classifies each internal node of a family tree in speciation and duplication events and assigns it to the phylogenetic level (or age class) in which these events are detected (Vilella *et al* 2009). We used this information in our pipeline to establish two alternative definitions of gene birth events. We defined *gene family age* as the last common ancestor to all the species containing a member of the gene family (i.e. the phylostratum defined by the root node of the gene family tree). We also defined the *gene age* as the phylostratum assigned to the last genomic event leading to the birth of an extant gene. So, the *gene age* of genes with a no-duplication origin (singletons) corresponds to their *gene family age*, while for a duplicated gene it represents the ancestral species where its last duplication event was detected. For this purpose, we only considered duplication events showing a consistency score (Vilella *et al* 2009) above 0.3. When this score was exactly 0, we considered that this duplication was an artifact of the phylogenetic reconstruction, and we ignored this node and established *gene age* using the previous nodes in the tree. Duplication nodes with consistencies between 0 and 3 were considered unclear and *gene age* could not be assigned.

**Yeast Gene birth dating**

For *Saccharomyces cerevisiae* we performed a *gene family* birth dating analysis based on PSI-BLAST homology searches, following a pipeline similar to that described recently (Domazet-Lošo *et al* 2007). For this we created a 3-rounds PSI-BLAST (Altschul *et al* 1997) profile for every yeast protein against a 90% sequence identity non-redundant version of UniProt database. We used these profiles to detect homologues for each yeast protein-coding gene by searching against sequence databases created from the UniProt database for a range of taxonomic divisions. Each database contained only those sequences from species with the same last common ancestor as *Saccharomyces cerevisiae*. In this way, detection of a significant hit (e-value < $10^{-5}$) in a given database implies that an ancestor was present in the corresponding ancestral species prior to the evolutionary split. We date the *gene family* birth event to the evolutionary time period (or phylostratum) represented by the most recent database in which we detected the presence of an ancestor gene.

**Database annotations**

Annotations were taken from a range of resources. The Protein Existence annotations came from UniProt (UniProt Consortium 2013), if there was more than one splice isoform the isoform with the highest ranked evidence was taken as the representative of the gene. We also downloaded all UniProt caution advice, three in particular were indicative of genes with little or no proteomics evidence, those that warned of "dubious CDS prediction", "pseudogene", preliminary data". UniProt also annotated a number of genes as "Obsolete".

UniProt also annotates proteins manually with protein evidence. Human proteins are particularly well annotated within UniProt. Protein evidence in UniProt is organised in five levels that are in order of decreasing evidence: "Protein", "Transcript", "Homology", "Predicted" and "Uncertain".

Ensembl gene descriptions were also a useful source of annotations. Ensembl gene descriptions allowed us to generate subsets of genes annotated as "pseudogene", "readthrough", "non-coding", "non-functional", "antisense" and "opposite strand".

We were also able to generate subsets of genes from GENCODE tags. GENCODE transcripts have three types of "status" tag, "KNOWN" is the most reliable, "PUTATIVE" identifies the transcripts with the least evidence. Where a gene had multiple splice variants we took the transcript with the highest ranked tag to represent the gene. GENCODE also has a "class" tag. Most transcripts are tagged as "protein coding", but there are some transcripts tagged as "nonsense_mediated_decay" (NMD). Where all gene transcripts were in the nonsense_mediated_decay class, we tagged the gene as NMD.

Finally the GENCODE project is manually annotating all transcripts with a transcription support level in collaboration with Ensembl. The annotation levels for multiple exon transcripts range from "mRNA covers all introns" (the highest annotation level) to "suspect ESTs" and "no evidence", the two lowest levels of transcript support. Once again for those genes with alternatively spliced transcripts, the highest ranked transcription support level was taken as the transcript support level for the whole gene.

**Feature selection**

A total of 19 features from a range of sources correlated with very low peptide detection rates. The list of features with very low peptide detection rates is shown below, ordered by the number of genes that has each of the features. The annotation source is in brackets.

**Class 1. Genes with no protein-like features (from APPRIS)**

These were genes that had no protein features and medium to high MI score (because we detected peptides for 33% of genes with no protein features but that have good conservation - MI scores lower than 0.4).

**Class 2. Genes with poor protein coding conservation (APPRIS)**

Here we included all genes that had an INERTIA MI score above 1 and those cases where INERTIA did not produce a score because few species had related sequences.

**Class 3. Primate genes (ENSEMBL Compara)**

This class included those genes with primate gene family age. We detected peptides for just 5 of the 563 genes annotated as appearing since primates.

**Class 4. PUTATIVE genes (GENCODE)**

These were genes which have all transcripts annotated as PUTATIVE by GENCODE. PUTATIVE transcripts are the least reliable level of GENCODE annotations.

**Classes 5, 6 and 7. Genes with weak Protein Evidence (UniProt)**

Genes where all splice isoforms were annotated with Homology evidence or worse had little evidence of protein expression (the best was Homology with a 6.87% peptide detection rate). The relation between protein evidence and detection can be seen in **supplementary figure 9**.

**Class 8. Genes with (semi-)circular annotation (UniProt/Ensembl)**

There were 336 genes where Ensembl took their description from a UniProt entry, and the corresponding UniProt entry linked back to Ensembl with the following caution: "*The sequence shown here is derived from an Ensembl automatic analysis pipeline and should be considered as preliminary data*". There was peptide evidence for just 4 of these genes.

**Classes 9 and 10. Genes with UniProt Cautions**

These were genes with other cautions in the UniProt annotations, either because the isoforms were tagged as potential pseudogenes or because they were tagged as dubious CDS predictions. There were 126 genes with these two cautions.

**Class 11. Obsolete genes (Ensembl/UniProt)**

130 genes had Ensembl descriptions that pointed to UniProt/TrEMBL protein entries tagged as "Obsolete". None of these genes had any evidence of protein coding. Most of them were no longer annotated in GENCODE 18.

**Class 12. Genes supported by suspect ESTs (GENCODE)**

There were 98 genes with transcripts supported only by "suspect ESTs". We did not detect peptides for any of these genes. The relation between transcript support and detection can be seen in **supplementary figure 10**)

**Class 13. Nonsense-mediated decay genes (GENCODE)**

GENCODE include transcripts annotated as nonsense-mediated decay targets within the protein coding set. There were 75 genes annotated solely with nonsense-mediated decay (NMD) transcripts.

**Class 14. Pseudogenes (Ensembl)**

These were genes that were tagged with the word "pseudogene" in the Ensembl description. There were 75 in all.

**Class 15. Read-through genes (Ensembl/GENCODE)**

There were 229 genes annotated as "read-through" in the Ensembl description or by GENCODE. We detected peptides for just two of these genes.

**Class 16. Non-functional genes (Ensembl)**

These were genes that are annotated as "non-functional" by Ensembl as part of their description field. Many of these were T-cell receptors and immunoglobulins. We did not detect peptides for any of these genes.

**Class 17. Non-coding genes (Ensembl)**

38 genes of the genes in GENCODE12 were tagged as "non-coding" in the Ensembl description field. As might be expected we did not detect peptides for any of these genes.

**Class 18. Antisense/opposite strand genes (Ensembl)**

Annotated as anti-sense or opposite strand as part of the Ensembl description. We did not detect peptides for any of these 25 genes.

**Class 19. Miscellaneous RNA (Ensembl)**

There were 7 genes tagged in the Ensembl description field as "long intergenic non-protein coding RNA" or "microRNA". Again we did not detect peptides for any of these genes.

**Reading Frame Conservation**

We calculated RFC scores for all protein coding genes in the GENCODE 12 annotation and for a set of non-coding genes. Alignments for the GENCODE12 transcripts were obtained from the UCSC 46-way mammalian multiple alignments (Lindblad-Toh *et al* 2011). Alignments for the non-coding regions were downloaded from the Ensembl Compara alignments (Flicek *et al* 2011). RFC scores were calculated from pairwise alignments between human and chimpanzee, human and macaque, human and dog, and human and mouse. The RFC score for each gene/non-coding region was calculated as the proportion of aligned bases that stay in frame.

In all cases we calculated RFC scores across three frames and took the highest scoring frame to avoid cases where misalignment at the 5' end skews the final score. For the GENCODE 12 annotations we took the APPRIS principal variant (Rodriguez *et al* 2013) as the representative for the gene. For the non-coding genes we took the longest transcript. The coding genes were split into three groups for comparison. Those 11,840 for which we detected peptides, the 1,940 genes from the potential non-coding set that we did not find peptides for and the remaining protein coding genes.

**Predicting gene type for the potential non-coding set**

We generated four sub-groups (read-throughs, possible coding, possible non-coding, possible pseudogenes) from the potential non-coding set. Read-through genes were the 229 read-through genes. Possible protein coding genes were those genes for which we detected peptides, genes that had protein evidence from other sources and genes that had not evolved recently, that had good conservation, and that had few atypical protein features. Possible pseudogenes were genes the annotated as pseudogenes by UniProt, Ensembl or the Clamp analysis, genes from highly duplicated families annotated as non-functional and genes that we felt were not non-protein coding and nevertheless had clear protein-like features. Possible non-coding genes were those annotated as non-coding by Ensembl or the Clamp analysis, and genes that had no clear protein features and were not conserved.

Several genes were left out of these sets, specifically genes from multiple recent duplications where there were several genes with practically identical scores, and genes read from frames opposite coding exons. There were 229 read-through genes, 343 possible pseudogenes, 968 possible non-coding genes and 392 possible protein coding genes.

**Acknowledgements**

This work was supported by the National Institutes of Health (NIH, grant number U41 HG007234) and by the Spanish Ministry of Science and Innovation (grant numbers BIO2007-666855, RD07-0067-0014, COMBIOMED). JMR is supported by the Spanish National Institute of Bioinformatics (www.inab.org), a platform of the "Instituto de Salud Carlos III". The authors would like to thank Daniel Rico for suggestions and discussions regarding gene expression and gene family age data.

**Author contributions**

IE performed the CNIO proteomics analysis; JMR performed the RFC analysis and provided the data from APPRIS; DJ collected the gene expression data and carried out the gene birth analyses for both human and yeast; MD provided the GENCODE transcript support data; AF provided data for the read-through genes; MLT, IE, and AV designed the study; MLT, IE, and DJ analysed the data; MLT wrote the paper; IE, AF, JLH, DJ, AV and JV contributed text and comments to the manuscript.

**Disclosure Declaration**

The authors have no conflicts of interest.

**Figure Legends**

**Figure 1. The percentage of genes identified in proteomics experiments as a function of gene conservation.**
Gene conservation is expressed using MI score, displayed in bins. Bin "0" is MI scores from 0 to 0.019, "0.02" is from 0.02 to 0.039, *etc*. The "missing" genes are those where the conservation was so poor that INERTIA was not able to generate a score.

**Figure 2. The percentage of genes for which peptides are detected in proteomics experiments against gene family age.**
Genes with gene families that appeared in the oldest phylogenetic divisions (towards the left) are identified much more often in proteomics experiments than those genes with families that appeared in the most recent phylogenetic divisions.

**Figure 3. Transcript ubiquity for human genes**
UniGene contains transcript evidence for most human genes over 45 different tissues. For each gene we counted the number of tissues in which there was transcript evidence of at least 5 or more transcripts per million. We separated the numbers of tissues in which transcripts were detected in UniGene into 10 bins and calculated the percentage of genes in each of the ten bins. We split the GENCODE12 genes into 3 groups, those genes for which we found peptides ("Detected" in dark red), those genes for which we did not find peptides that were also in the potential non coding set ("Potential NC" genes marked in yellow), and those for which we did not find peptides, but that were not in the potential non-coding set ("Not Detected" genes, in orange).

**Figure 4. RFC scores for pairwise alignments with 4 species**
The RFC scores are calculated as per the methods section. RFC scores for alignments between (A) human and chimp, (B) human and macaque, (C) human and mouse and (D) human and dog. We split the GENCODE12 genes into 3 groups, those genes for which we found peptides ("Detected" in dark red), those genes for which we did not find peptides and that were also in the potential non coding set ("Potential NC" genes marked in yellow), and those for which we did not find peptides, but that were not in the potential non-coding set ("Not Detected" genes, in orange). As a comparison we included the results for a set of long non-coding genes ("Non-coding" shown in blue). RFC scores are shown on the y-axis, the x-axis in all the figures is the proportion of the RFC scores we could calculate for each set. RFC scores are ordered from highest to lowest.

**Figure 5. Genes in the potential non-coding set for which we find evidence of peptides from other sources.**
We searched multiple proteomics experiments and four different sources of protein evidence for evidence of the expression of proteins for all 2,001 genes in the potential non-coding set. Here we show the numbers of genes detected by each source and the overlap. The 7 proteomics studies are show in red, the Passel Database in green, the Human Protein Atlas in yellow and UniProt referenced papers in blue.

**Figures**

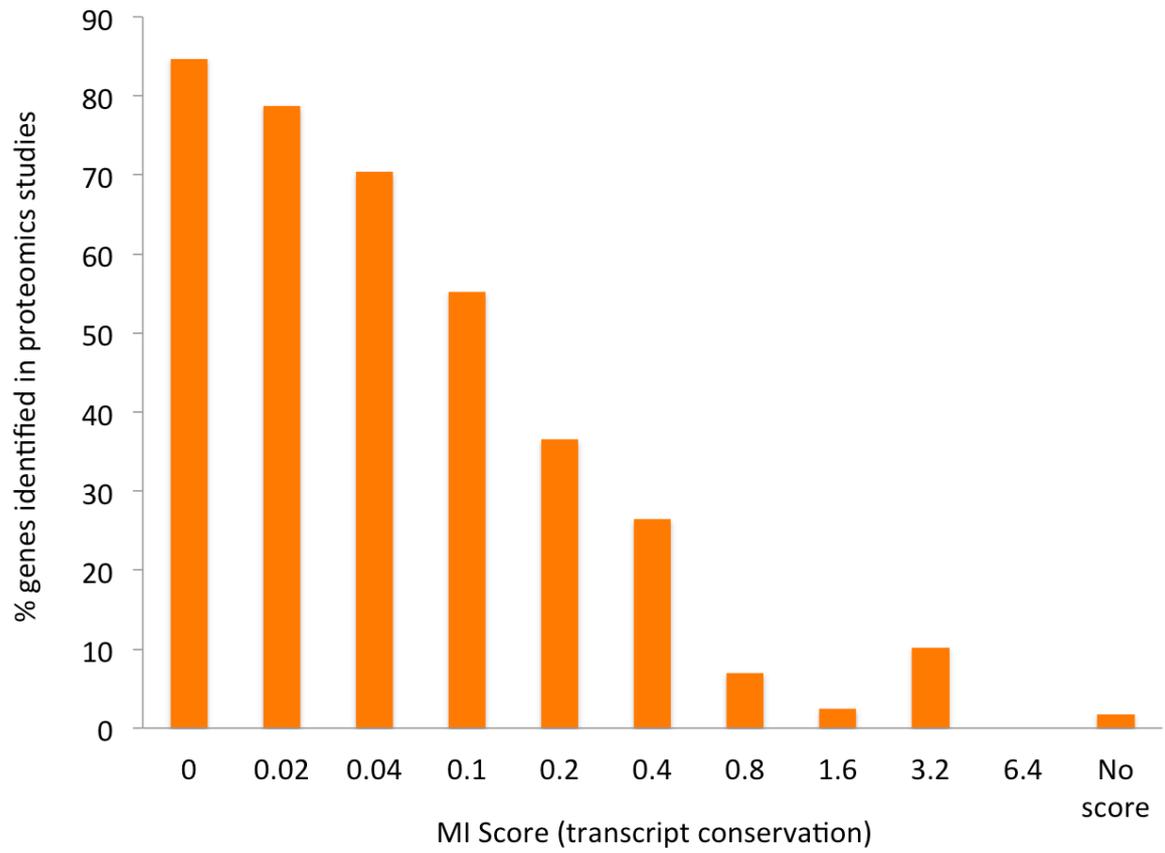

**Figure 1**

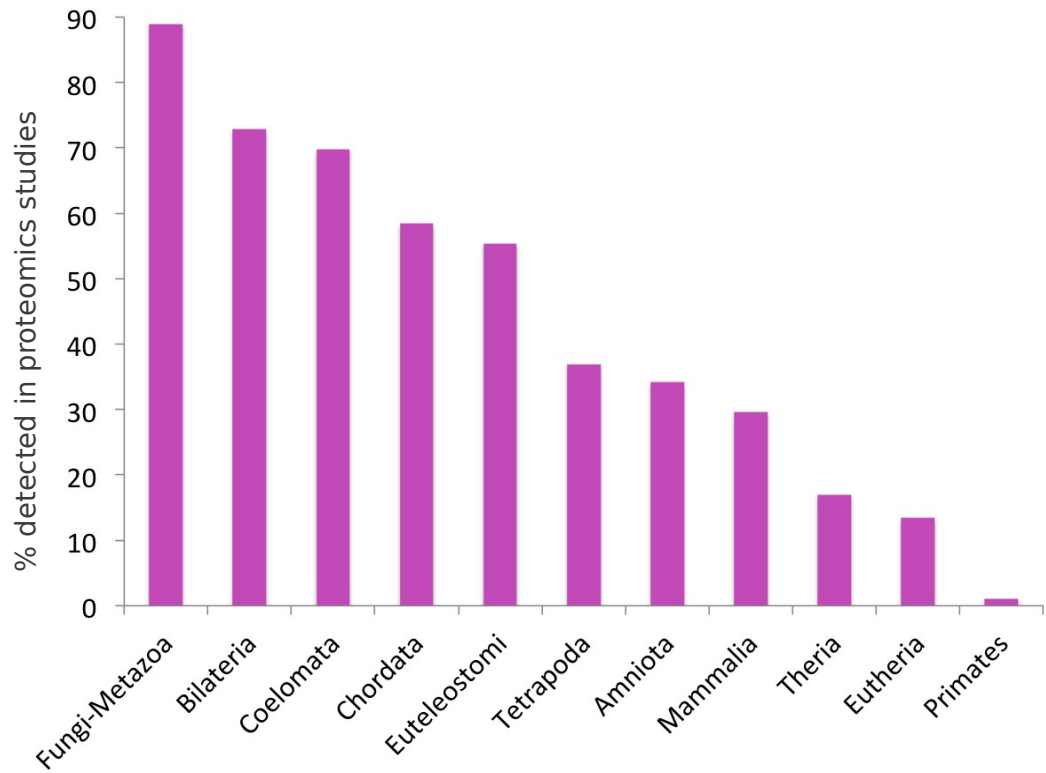

**Figure 2**

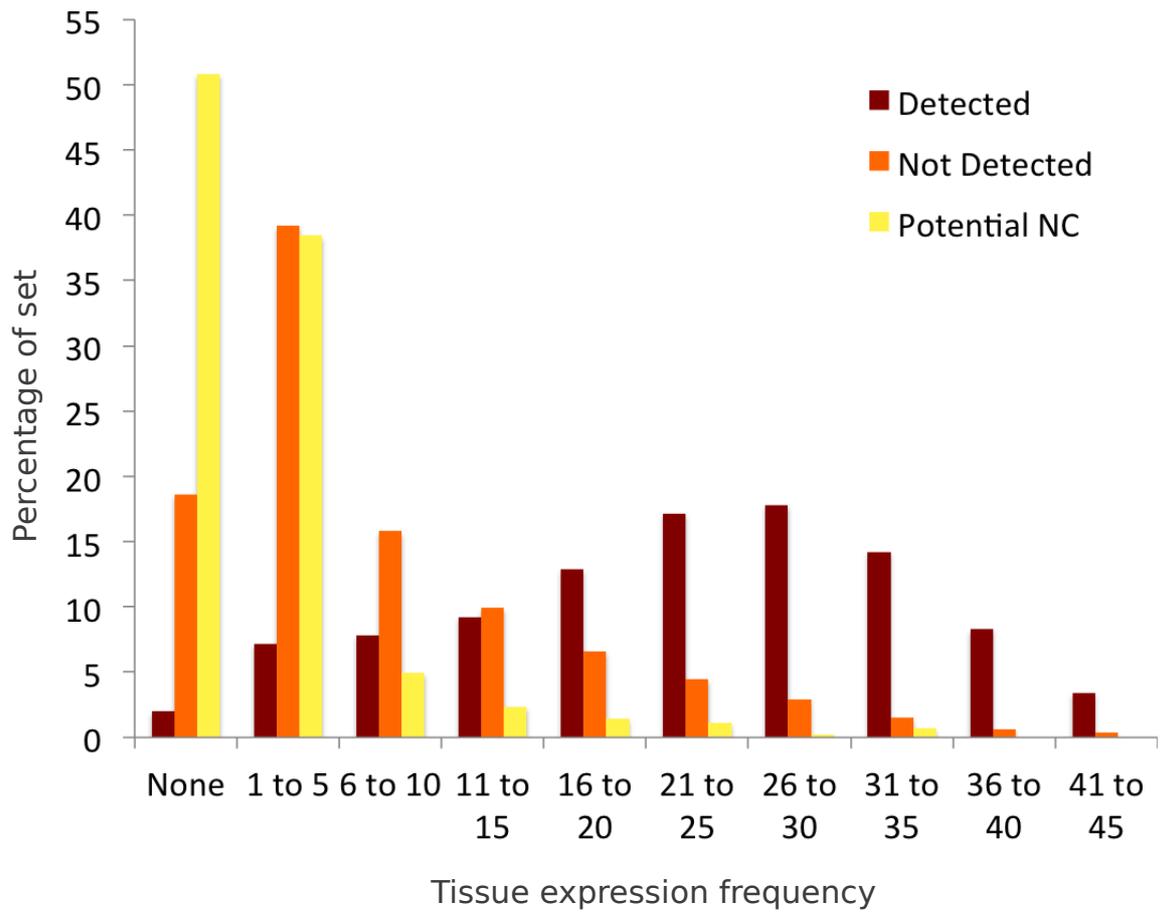

**Figure 3**

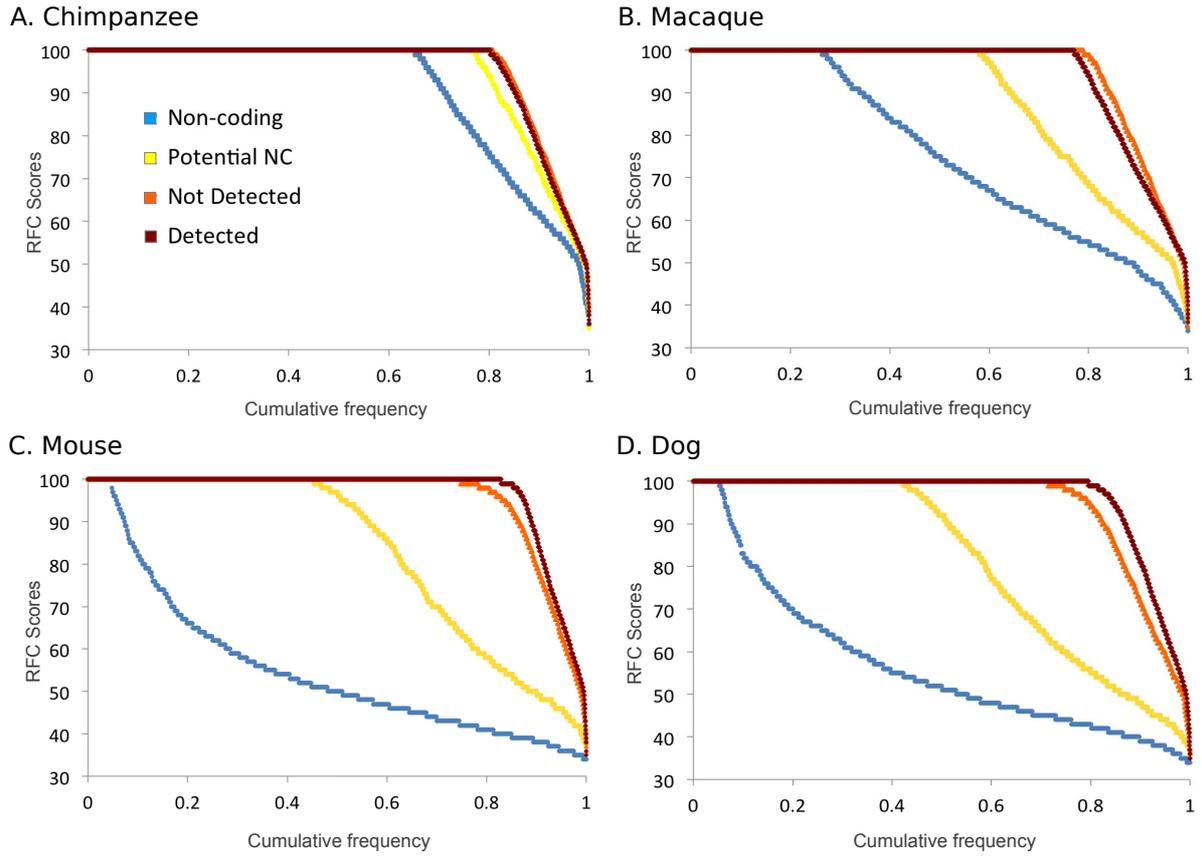

Figure 4

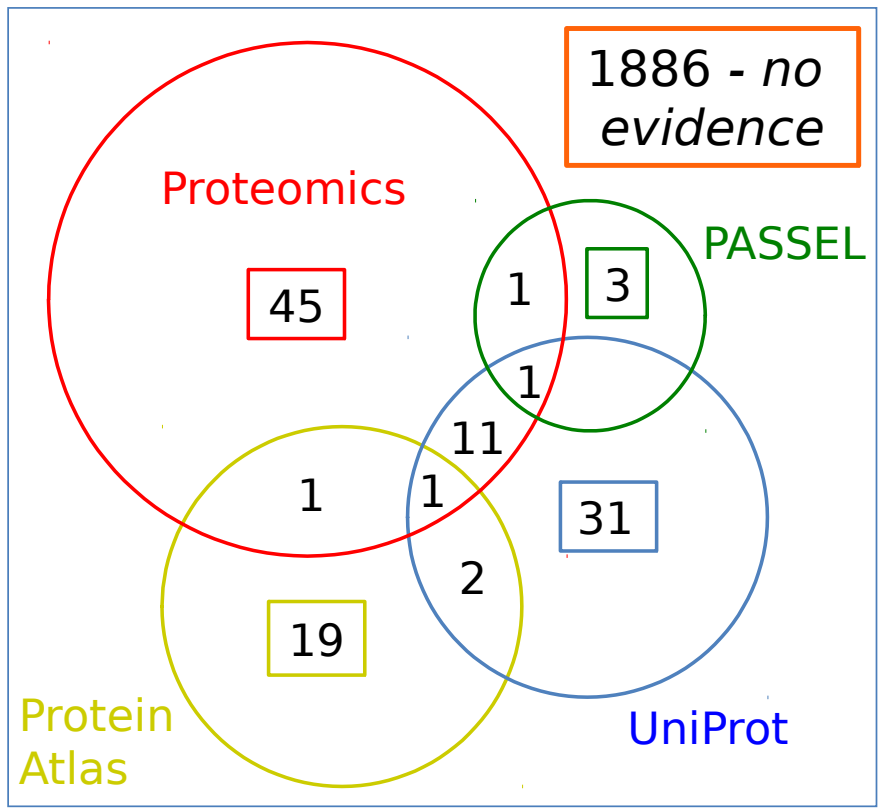

Figure 5

**Tables**

| Features | Genes | Peptide Detection |
| --- | ---: | ---: |
| Homology existence **[UP]** | 131 | **6.87%** |
| Pseudogene **[E]** | 75 | **6.67%** |
| PUTATIVE transcripts **[G]** | 434 | **2.53%** |
| Caution – pseudogene **[UP]** | 79 | **2.53%** |
| Caution – dubious CDS **[UP]** | 47 | **2.13%** |
| Poor conservation (MI score) **[A]** | 987 | **2.03%** |
| Predicted existence **[UP]** | 507 | **1.58%** |
| No protein features **[A]** | 1212 | **1.32%** |
| Nonsense mediated decay **[G]** | 78 | **1.28%** |
| Circular annotation **[E/UP]** | 336 | **1.19%** |
| Uncertain existence **[UP]** | 100 | **1.00%** |
| Primate gene family **[E]** | 563 | **0.89%** |
| Read-through **[E/G]** | 229 | **0.87%** |
| Obsolete **[E/UP]** | 130 | **0.00%** |
| Dubious EST support **[E/G]** | 98 | **0.00%** |
| Non-functional **[E]** | 44 | **0.00%** |
| Non-coding **[E]** | 38 | **0.00%** |
| Antisense/Opposite Strand **[E]** | 25 | **0.00%** |
| Miscellaneous RNA **[E]** | 7 | **0.00%** |

**Table 1. The 19 features used to select the potential non-coding set**

Each feature is explained in more detail in the method and supplementary sections. The source of each feature is indicated in square brackets (A=APPRIS, E=Ensembl, G=GENCODE, UP=UniProt). For each feature we also show the number of genes with the feature and the proportion that we identify in the 7 datasets.

# References

Aebersold R, Mann M. 2003. Mass spectrometry-based proteomics *Nature* **422**: 198– 207.

Altschul SF, Madden TL, Schäffer AA, Zhang J, Zhang Z, Miller W, Lipman DJ. 1997. Gapped BLAST and PSI-BLAST: a new generation of protein database search programs. *Nucleic Acids Res*. **25**: 3389-3402.

Beck, M, Schmidt A, Malmstroem J, Claassen M, Ori A, Szymborska A, Herzog F, Rinner O, Ellenberg J, Aebersold R. 2011. The quantitative proteome of a human cell line. *Mol Syst Biol.* **7**: 549.

Blanchette M, Kent WJ, Riemer C, Elnitski L, Smit AF, Roskin KM, Baertsch R, Rosenbloom K, Clawson H, Green ED, *et al*. 2004. Aligning multiple genomic sequences with the threaded blockset aligner. *Genome Res.* **14**: 708-715.

Brosch M, Saunders GI, Frankish A, Collins MO, Yu L, Wright J, Verstraten R, Adams DJ, Harrow J, Choudhary JS, *et al.* 2011. Shotgun proteomics aids discovery of novel protein-coding genes, alternative splicing, and "resurrected" pseudogenes in the mouse genome. *Genome Res.* **21**: 756-767.

Buljan M, Frankish A, Bateman A. 2010. Quantifying the mechanisms of domain gain in animal proteins. *Genome Biol.* **11**: R74.

Cannarozzi G, Schneider A, Gonnet G. 2007. A phylogenomic study of human, dog, and mouse. *PLoS Comp Biol.* **3**: e2.

Clamp M, Fry B, Kamal M, Xie X, Cuff J, Lin MF, Kellis M, Lindblad-Toh K, Lander ES. 2007. Distinguishing protein-coding and noncoding genes in the human genome. *Proc Natl Acad Sci USA* **104**: 19428-19433.

Colaert N, Van Huele C, Degroeve S, Staes A, Vandekerckhove J, Gevaert K, Martens L. 2011. Combining quantitative proteomics data processing workflows for greater sensitivity. *Nat Methods* **8**: 481-483.

Cox J, Neuhauser N, Michalski A, Scheltema RA, Olsen JV, Mann M. 2011. Andromeda: a peptide search engine integrated into the MaxQuant environment. *J Proteome.* **10**: 1794-1805.

Craig R, Beavis RC. 2003. A method for reducing the time required to match protein sequences with tandem mass spectra. *Rapid Commun. Mass Sp.* **17**: 2310-2316.

Craig R, Cortens JP, Beavis RC. 2004. Open source system for analyzing, validating, and storing protein identification data. *J Proteome Res.* **3**: 1234-1242.

Domazet-Lošo T, Brajković J, Tautz D. 2007. A phylostratigraphy approach to uncover the genomic history of major adaptations in metazoan lineages. *Trends Genet.* **23**: 533–539.

Emanuelsson O, Brunak S, von Heijne G, Nielsen H. 2007. Locating proteins in the cell using TargetP, SignalP and related tools. *Nat Protoc.* **2**: 953-971.

ENCODE Project Consortium. 2007. Identification and analysis of functional elements in 1% of the human genome by the ENCODE pilot project. *Nature* **447**: 799-816.

ENCODE Project Consortium. 2012. An integrated encyclopedia of DNA elements in the human genome. *Nature* **489**: 57-74.

Ezkurdia I, del Pozo A, Frankish A, Rodriguez JM, Harrow J, Ashman K, Valencia A, Tress ML. 2012. Comparative proteomics reveals a significant bias toward alternative protein isoforms with conserved structure and function. *Mol Biol Evol.* **29**: 2265-2283.

Farrah T, Deutsch EW, Kreisberg R, Sun Z, Campbell DS, Mendoza L, Kusebauch U, Brusniak MY, Hüttenhain R, Schiess R, *et al*. 2012. PASSEL:the PeptideAtlas SRMexperiment library. *Proteomics* **12**: 1170-1175.

Farrah T, Deutsch EW, Hoopmann MR, Hallows JL, Sun Z, Huang CY, Moritz RL. 2013. The state of the human proteome in 2012 as viewed through PeptideAtlas. *J Proteome Res.* **12**: 162-171.

Flicek P, Amode MR, Barrell D, Beal K, Brent S, Chen Y, Clapham P, Coates G, Fairley S, Fitzgerald S *et al*. 2011. Ensembl 2011. *Nucleic Acids Res.* **39**: D800-D806.

Flicek P, Ahmed I, Amode MR, Barrell D, Beal K, Brent S, Carvalho-Silva D, Clapham P, Coates G, Fairley S *et al*. 2013. Ensembl 2013. *Nucleic Acids Res.* **41**: D48-D55.

Geiger T, Wehner A, Schaab C, Cox J, Mann M. 2012. Comparative proteomic analysis of eleven common cell lines reveals ubiquitous but varying expression of most proteins. *Mol Cell Proteomics* **11**: M111 014050.

Guigó R, Flicek P, Abril JF, Reymond A, Lagarde J, Denoeud F, Antonarakis S, Ashburner M, Bajic VB, Birney E, *et al*. 2006. EGASP: the human ENCODE Genome Annotation Assessment Project *Genome Biol.* **7**: S2.

Harrow J, Denoeud F, Frankish A, Reymond A, Chen CK, Chrast J, Lagarde J, Gilbert JG, Storey R, Swarbreck D, *et al*. 2006. GENCODE: producing a reference annotation for ENCODE. *Genome Biol.* **7**: S4.

Harrow J, Frankish A, Gonzalez JM, Tapanari E, Diekhans M, Kokocinski F, Aken BL, Barrell D, Zadissa A, Searle S, *et al*. 2012. GENCODE: the reference human genome annotation for The ENCODE Project. *Genome Res.* **22**: 760-774.

Huang da W, Sherman BT, Lempicki, RA. 2009. Systematic and integrative analysis of large gene lists using DAVID bioinformatics resources. *Nat Protoc.* **4**: 44-57.

Hubbard T, Barker D, Birney E, Cameron G, Chen Y, Clark L, Cox T, Cuff J, Curwen V, Down T, *et al.* 2007. The Ensembl genome database project. *Nucleic Acids Res.* **30**: 38-41.

Huerta-Cepas J, Dopazo H, Dopazo J, Gabaldón T. 2007. The human phylome. *Genome Biol.* **8**: R109.

International Human Genome Sequencing Consortium. 2001. Initial sequencing and analysis of the human genome. *Nature* **409**: 860−921.

International Human Genome Sequencing Consortium. 2004. Finishing the euchromatic sequence of the human genome. *Nature* **431**: 931–945.

Jones DT. 2007. Improving the accuracy of transmembrane protein topology prediction using evolutionary information. *Bioinformatics* **23**: 538-544.

Käll L, Krogh A, Sonnhammer EL. 2004. A combined transmembrane topology and signal peptide prediction method. *J Mol Biol.* **338**: 1027-1036.

Koenig T, Menze BH, Kirchner M, Monigatti F, Parker KC, Patterson T, Steen JJ, Hamprecht FA, Steen H. 2008. Robust prediction of the MASCOT score for an improved quality assessment in mass spectrometric proteomics. *J Proteome Res.* **7**: 3708–3717.

Kristensen AR, Gsponer J, Foster LJ. 2013. Protein synthesis rate is the predominant regulator of protein expression during differentiation. *Mol Syst Biol.* **9**: 689.

Lassmann T, Sonnhammer EL. Kalign - an accurate and fast multiple sequence alignment algorithm. *BMC Bioinformatics,* **6**: 298.

Lee HJ, Jeong SK, Na K, Lee MJ, Lee SH, Lim JS, Cha HJ, Cho JY, Kwon JY, Kim H, *et al*. 2013. Comprehensive genome-wide proteomic analysis of human placental tissue for the Chromosome-Centric Human Proteome Project. *J Proteome Res.* **12**: 2458-2466.

Lindblad-Toh K, Garber M, Zuk O, Lin MF, Parker BJ, Washietl S, Kheradpour P, Ernst J, Jordan G, Mauceli E, *et al*. 2011. A high-resolution map of human evolutionary constraint using 29 mammals. *Nature* **478**: 476-482.

Lopez G, Maietta P, Rodriguez J-M, Valencia A, Tress ML. 2011. firestar--advances in the prediction of functionally important residues. *Nucleic Acids Res.* **39**: W235-W241.

Löytynoja A, Goldman N. 2005. An algorithm for progressive multiple alignment of sequences with insertions. *Proc Natl Acad Sci USA* **102**: 10557-10562.

Mallick P, Kuster B. 2010. Proteomics: a pragmatic perspective *Nat Biotechnol.* **28:** 695–709.

Massingham T, Goldman N. 2005. Detecting amino acid sites under positive selection and purifying selection. *Genetics* **169**: 1753-1762.

Moore R, Young M, Lee T. 2002. Qscore: an algorithm for evaluating SEQUEST database search results. *J Am Soc Mass Spectrom.* **13**: 378-386.

Munoz J, Low TY, Kok YJ, Chin A, Frese CK, Ding V, Choo A, Heck AJ. 2011. The quantitative proteomes of human- induced pluripotent stem cells and embryonic stem cells. *Mol Syst Biol.* **7**: 550.

Nagaraj N, Wisniewski JR, Geiger T, Cox J, Kircher M, Kelso J, Pääbo S, Mann M. 2011. Deep proteome and transcriptome mapping of a human cancer cell line. *Mol Syst Biol.* **7**: 548.

Neuhauser N, Nagaraj N, McHardy P, Zanivan S, Scheltema R, Cox J, Mann M. 2013. High performance computational analysis of large-scale proteome datasets to assess incremental contribution to coverage of the human genome. *J. Proteome Res.* **12**: 2858-2868.

Pennisi E. 2003. A low gene number wins the GeneSweep pool. *Science* **300**: 1484.

Picotti P, Clément-Ziza M, Lam H, Campbell DS, Schmidt A, Deutsch EW, Röst H, Sun Z, Rinner O, Reiter L, *et al*. 2013. A complete mass-spectrometric map of the yeast proteome applied to quantitative trait analysis. *Nature* **494**: 266-270.

Punta M, Coggill PC, Eberhardt RY, Mistry J, Tate J, Boursnell C, Pang N, Forslund K, Ceric G, Clements J, *et al*. 2012. The Pfam protein families database. *Nucleic Acids Res*. **40**: D290-D301.

Rodriguez JM, Maietta P, Ezkurdia I, Pietrelli A, Wesselink JJ, Lopez G, Valencia A, Tress ML. 2013. APPRIS: annotation of principal and alternative splice isoforms. *Nucleic Acids Res.* **41**: D110-D117.

Rose PW, Beran B, Bi C, Bluhm WF, Dimitropoulos D, Goodsell DS, Prlic A, Quesada M, Quinn GB, Westbrook JD, *et al*. 2011. The RCSB Protein Data Bank: redesigned web site and web services. *Nucleic Acids Res.* **39**: 392-401.

Roux J, Robinson-Rechavi M. 2011. Age-dependent gain of alternative splice forms and biased duplication explain the relation between splicing and duplication. *Genome Res.* **21**: 357–363.

Tanner S, Shen Z, Ng J, Florea L, Guigó R, Briggs SP, Bafna V. 2007. Improving gene annotation using peptide mass spectrometry. *Genome Res.* **17**: 231-239.

Uhlen M, Oksvold P, Fagerberg L, Lundberg E, Jonasson K, Forsberg M, Zwahlen M, Kampf C, Wester K, Hober S, *et al*. 2010. Towards a knowledge-based Human Protein Atlas. *Nat Biotechnol.* **28**: 1248-1250.

UniProt Consortium. 2013. Update on activities at the Universal Protein Resource (UniProt) in 2013. *Nucleic Acids Res.* **41**: D43-D47.

Veeramah KR, Thomas MG, Weale ME, Zeitlyn D, Tarekegn A, Bekele E, Mendell NR, Shephard EA, Bradman N, Phillips IR. 2008. The potentially deleterious functional variant flavin-containing monooxygenase 2*1 is at high frequency throughout sub-Saharan Africa. *Pharmacogenet. Genomics* **18**: 877-886.

Venter JC, Adams MD, Myers EW, Li PW, Mural RJ, Sutton GG, Smith HO, Yandell M, Evans CA, Holt RA *et al.* 2001. The sequence of the human genome. *Science* **291**: 1304−1351.

Viklund H, Elofsson A. 2004. Best alpha-helical transmembrane protein topology predictions are achieved using hidden Markov models and evolutionary information. *Protein Sci.* **13**: 1908-1917.

Vilella AJ, Severin J, Ureta-Vidal A, Heng L, Durbin R, Birney E. 2009. EnsemblCompara GeneTrees: Complete, duplication-aware phylogenetic trees in vertebrates. *Genome Res.* **19**: 327–335.

Wheeler DL, Church DM, Federhen S, Lash AE, Madden TL, Pontius JU, Schuler GD, Schriml LM, Sequeira E, Tatusova TA, *et al.* 2003. Database resources of the National Center for Biotechnology. *Nucleic Acids Res.* **31**: 28.

**Title:** The shrinking human protein coding complement: are there now fewer than 20,000 genes?

**Authors:** Iakes Ezkurdia[1], David Juan[2], Jose Manuel Rodriguez[3], Adam Frankish[4], Mark Diekhans[5], Jennifer Harrow[4], Jesus Vazquez[6], Alfonso Valencia[2,3], Michael L. Tress[2,*].

**Affiliations:**
1. Unidad de Proteómica, Centro Nacional de Investigaciones Cardiovasculares, CNIC, Melchor Fernández Almagro, 3, 28029, Madrid, Spain
2. Structural Biology and Bioinformatics Programme, Spanish National Cancer Research Centre (CNIO), Melchor Fernández Almagro, 3, 28029, Madrid, Spain
3. National Bioinformatics Institute (INB), Spanish National Cancer Research Centre (CNIO), Melchor Fernández Almagro, 3, 28029, Madrid, Spain
4. Wellcome Trust Sanger Institute, Wellcome Trust Campus, Hinxton, Cambridge CB10 1SA, UK
5. Center for Biomolecular Science and Engineering, School of Engineering, University of California Santa Cruz (UCSC), 1156 High Street, Santa Cruz, CA 95064, USA
6. Laboratorio de Proteómica Cardiovascular, Centro Nacional de Investigaciones Cardiovasculares, CNIC, Melchor Fernández Almagro, 3, 28029, Madrid, Spain

**The potential non-coding set, expanded version**

**Class 1. Genes without protein-like features (data from APPRIS)**

This is the largest group of genes and in some ways the most difficult to explain – just because a gene does not have any measurable protein features does not mean that we would expect to find no peptides. These genes do not have similarity to known 3D structures; so many (perhaps most) of these genes are likely to have long disordered regions. However, disorder alone cannot explain why we do not detect proteins for genes without protein features – we detect peptides for 65% of those genes that are predicted to have greater than 50 % disordered residues.

We limited the genes in this set to those with MI scores over 0.4 – we detected peptides for 33% of genes with no protein features but MI scores lower than 0.4 – and we detected peptides for just 16 of the 1212 genes in this set. Three of the genes we detected peptides for were annotated as "small, proline-rich proteins".

**Class 2. Genes with poor protein coding conservation (APPRIS)**

Here we included all genes that had an INERTIA MI score above 1, along with those genes where INERTIA did not produce any score because there were too few species with orthologues or because all the exons were shorter than 42 bases. Many of the genes in this set have more than one feature that correlates with lack of peptide detection – over 700 of the genes with poor conservation also had no protein features, the ancestors of half those genes with measurable "family" age appeared during the development of primates. We detected peptides for just 18 of the 987 genes in this group. Three of the genes with most peptides were immunoglobulin genes.

**Class 3. Primate genes (Compara)**

We included those genes that had evolved since primates because we detected peptides for just 5 of the 563 genes annotated as appearing since primates. Again there is much overlap between the features; all but 18 of the genes in this group have at least one other feature that correlates with low peptide detection. We detected peptides for four of these 18 genes and there were UniProt paper references for the protein existence of four more, suggesting that some primate genes have evolved to be protein coding. Curiously four of the five proteins in this set for which we did detect peptides are annotated as being secreted by the cell (*HTN1*, *STATH*, *SEMG1*, and *SEMG2*). *HTN1* was also among the ten orphans with peptide evidence found in the Clamp study (**1**).

**Class 4. PUTATIVE genes (GENCODE)**

These are genes for which all transcripts are annotated as PUTATIVE by the GENCODE annotators. PUTATIVE transcripts are the least reliable level of annotation. We detected peptides for 11 of 434 genes annotated solely with PUTATIVE transcripts. Several genes (such as *ZBED5*, *C12orf75*, *SPA17* and *HOXD13*) with all transcripts annotated as PUTATIVE, but no other non-coding features are likely to be real protein coding genes that will in time have transcripts that are annotated as KNOWN. The second lowest level of GENCODE annotation is NOVEL and we only detect peptides for just 18% of genes that have all transcripts annotated as NOVEL or PUTATIVE.

**Classes 5, 6 and 7. Genes with weak Protein Evidence (UniProt)**
Three of the five levels of protein evidence defined by the UniProt curators were related with a lack of peptide detection. We detected peptides for a single gene out of the 100 annotated with the lowest level of reliability, "Uncertain", peptides for 8 genes of the 507 genes whose existence was "Predicted" and 9 genes of the 131 whose existence was backed by the "Homology" annotation. All genes annotated as Uncertain are also annotated with a UniProt "Caution", as are most genes annotated as "Predicted". Most genes annotated as "Predicted" also have no protein features and more than half of the genes annotated as "Predicted" by UniProt also have all their transcripts annotated as "PUTATIVE" by GENCODE. We detect peptides for just under 28% of genes annotated by UniProt with "transcript" protein evidence.

**Class 8. Genes with (semi-)circular annotation (UniProt/Ensembl)**
There are 336 genes annotated by Ensembl as having descriptions taken from UniProt/TrEMBL entries that in turn are annotated with the following caution: "*The sequence shown here is derived from an Ensembl automatic analysis pipeline and should be considered as preliminary data*". This is not completely circular logic because Ensembl are taking the protein description from UniProt, not the annotation. However, both databases appear to be pointing to each other as the source of the evidence, suggesting that there is little evidence for the expression of this gene. Most of these genes are annotated as "Predicted" by UniProt and have no protein-like features. Many of these genes are read-through genes. Less than a third have all their introns supported by a single mRNA according to GENCODE. There was peptide evidence for just 4 of these genes, all four genes appeared before the Euteleostomi division.

**Classes 9 and 10. Genes with other UniProt Cautions (UniProt)**
Other than the Ensembl automatic analysis pipeline warning there were two main reasons for a UniProt caution, either because the gene was a "*Product of a dubious CDS prediction*" or because the gene "*Could be the product of a pseudogene*". There were

126 genes with these cautions, 79 annotated as potential pseudogenes and 47 as dubious CDS. We detect peptides for two potential pseudogenes (including *WASH4P*, annotated as a pseudogene in the Ensembl description too, for which we detect peptides in five of the seven analyses) and a single gene with a dubious CDS, *TSPO*, for which we actually also detected evidence of alternative splicing.

**Class 11. Obsolete genes (Ensembl/UniProt)**

A total of 130 genes annotated by Ensembl with descriptions that lead to "Obsolete" UniProt/TrEMBL protein entries. More than half the genes are annotated with putative transcripts only by GENCODE and less than 30% have mRNA evidence that covers all the introns. More than half of the genes first appeared in primates and almost all of them have no protein features. None of the genes had any evidence of protein coding, so it looks as if the protein entries were retired with good reason. Most of these genes had been removed by the GENCODE v18 annotation.

**Class 12. Genes supported by suspect ESTs (GENCODE)**

GENCODE have classified all transcripts by their supporting evidence. We took the transcript with the best supporting evidence as the representative for the gene. There were 98 genes that were annotated with transcripts supported only by suspect ESTs. We did not detect peptides for any of these genes. Almost half these genes did not have any distinguishing tryptic peptides. For 87% of the genes where we were able to measure gene age, the gene seemed to be the product of a Human or HomoPanGorilla duplication.

**Class 13. Nonsense-mediated decay genes (GENCODE)**

GENCODE include transcripts annotated as nonsense-mediated decay within their protein coding set. There were 78 genes annotated solely with nonsense-mediated decay (NMD) transcripts. We detect peptide evidence for just one protein. We detect the bare minimum two peptides over the seven analyses for ENSG00000258539, a read-through gene that in any case should probably be annotated as a splice variant of *METTL10*. All but two of the genes annotated with NMD transcripts are read-through genes. *ACTN3*, a known protein-coding gene for which we have detected peptides in other experiments is one exception. It is currently tagged in GENCODE as a polymorphic pseudogene.

**14. Pseudogenes (Ensembl)**

These are genes that are tagged with the word "pseudogene" in the Ensembl description. There are 75 in all and we detect peptides for five, including *WASH4P* as previously mentioned. We also detect unique peptides for *HIST2H3PS2* in four analyses, but just two peptides each for the other two possible pseudogenes. Unlike the majority of the other groups with few detected peptides, the pseudogenes tend not to overlap with any of the other features – they are generally well conserved and have protein-like features. However, 9 are supported by suspect ESTs and another 13 have no transcript support at all. Fifteen of these pseudogenes are olfactory receptors, immunoglubulins or T-cell receptors.

### 15. Read-through genes (Ensembl/GENCODE)
There are 52 genes annotated as "read-through" in the Ensembl description and a further 177 that we detected within the GENCODE 12 annotation. It is particularly difficult to detect peptides for read-throughs because only those peptides that map to the sequence linking the two fused genes will map uniquely to the read-through protein. Indeed, a number of these read-through genes do not have a single unique distinguishing tryptic peptide. Even if part of the reason that we do not identify read-throughs in proteomics experiments is because there are few distinguishing peptides, read-throughs should probably not be annotated as separate genes at all. They are probably best described as alternative splice variants of one of the two genes. We detect peptides for just two of these genes. *PALM2-AKAP2*, for which we detect a single peptide in four different analyses, has sufficient supporting evidence to suggest that it may code for a protein. The vast majority of read-through genes are not likely to be protein coding due to their nature and the large (and increasing) numbers of read-through genes and transcripts hampers efforts to detect protein-coding genes – if read-through genes overlap other genes completely (a common occurrence) it is impossible to distinguish which gene peptides belong to.

### 16. Non-functional genes (Ensembl)
These are 44 genes that are annotated as "non-functional" by Ensembl as part of the description. Many of these are T-cell receptors and immunoglobulins. We do not detect peptides for any of these genes.

### 17. Non-coding genes (Ensembl)
These are 38 genes that are annotated as "non-coding" by Ensembl in their description. Many of these genes are also annotated as anti-sense or read-through. We do not detect peptides for any of these genes.

### 18. Antisense/opposite strand genes (Ensembl)
Annotated as anti-sense or opposite strand as part of the Ensembl description. We do not detect peptides for any of these 25 genes.

### 19. Miscellaneous RNA (Ensembl)

Seven genes are described as by Ensembl as "long intergenic non-protein coding RNA" or "microRNA".

**Supplementary Figures**

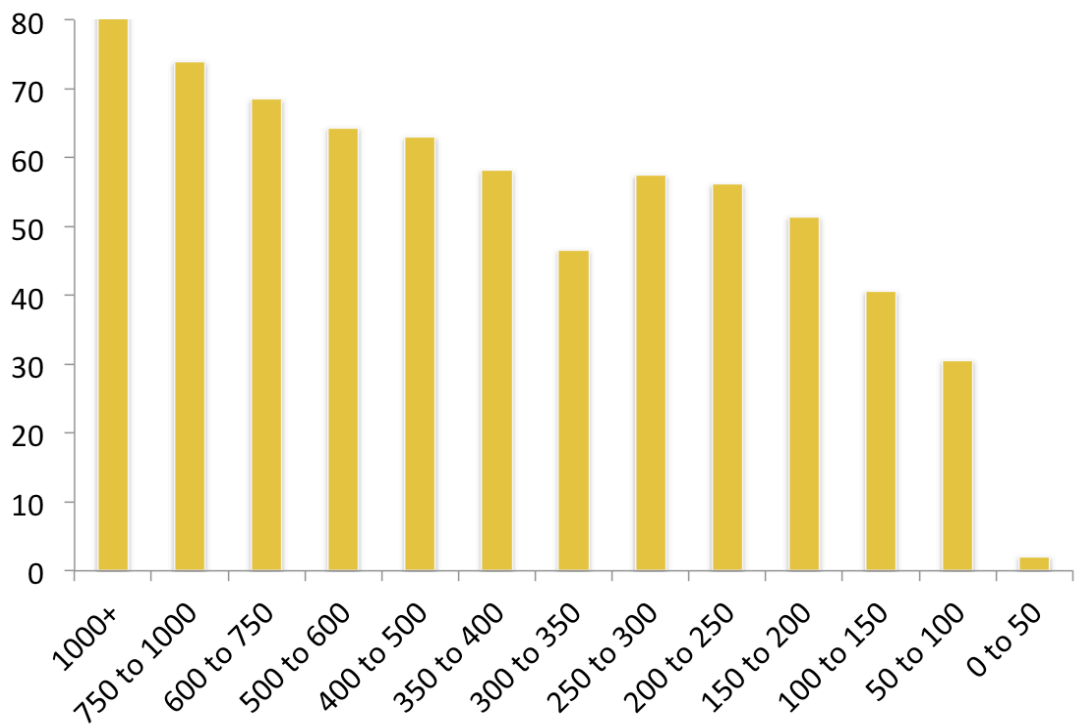

**Figure 1. The lengths of genes detected in proteomics analyses**
The percentage of genes identified in the proteomics studies against the length of the principal splice isoform from APPRIS for each gene.

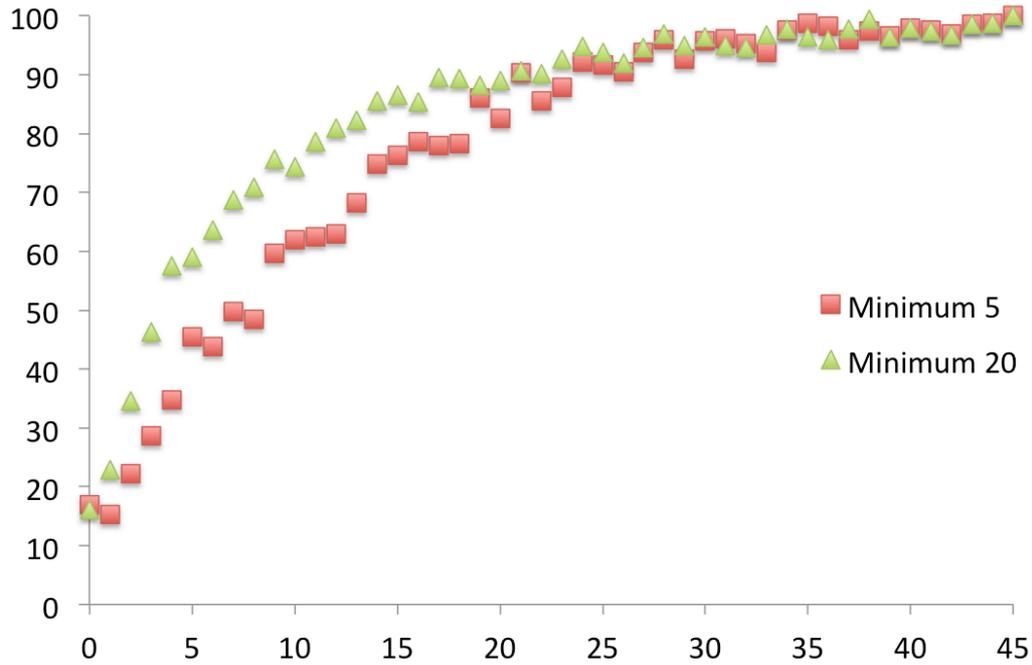

**Figure 2. The relationship between tissue expression and proteomics detection.**
The percentage of genes identified in the proteomics studies (y-axis) against the number of tissues in which transcript expression was found (x-axis). We used data from UniGene to look at ubiquity of transcript expression over a range of tissues. For each gene we counted the number of tissues in which there was transcript evidence of at least *5 or more transcripts per million.* Genes were binned according to the number of tissue types in which there was evidence of transcript expression.

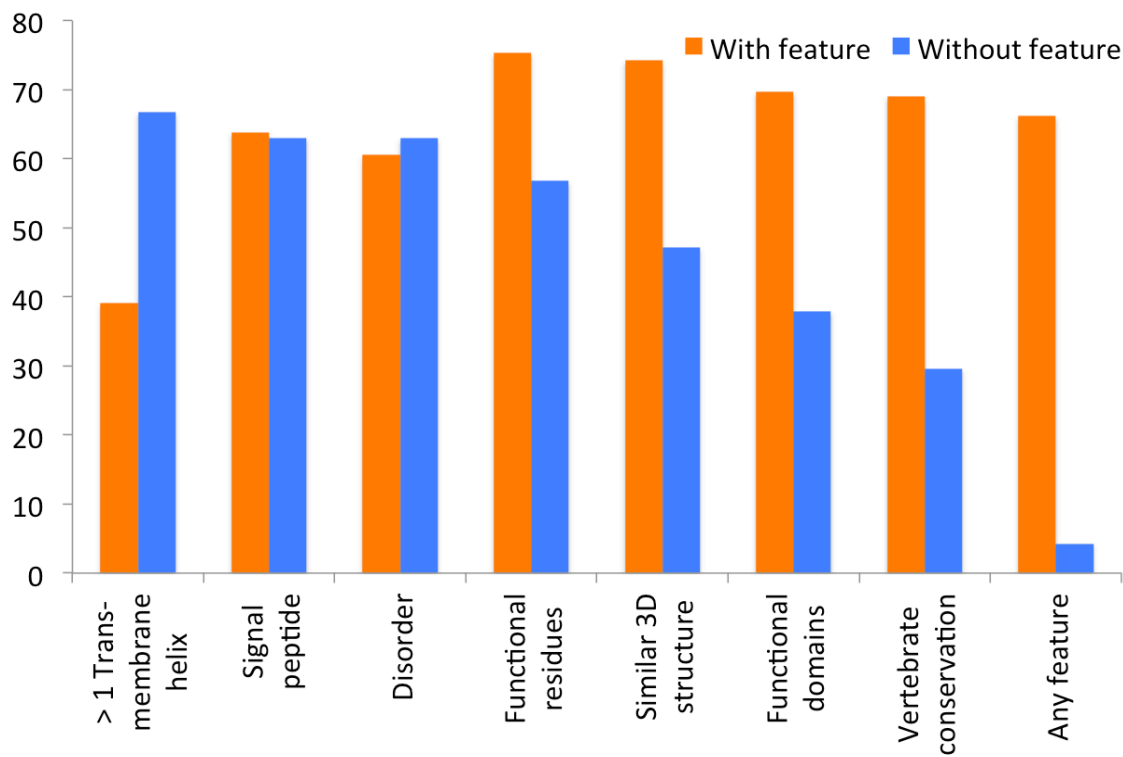

**Figure 3. The relationship between protein features and peptide identification**
We calculate the percentage of genes identified in the seven datasets (y-axis) for the presence or absence of a range of APPRIS features, for protein disorder and for those genes that were annotated with any single protein-like feature from APPRIS (the "Any feature" columns).

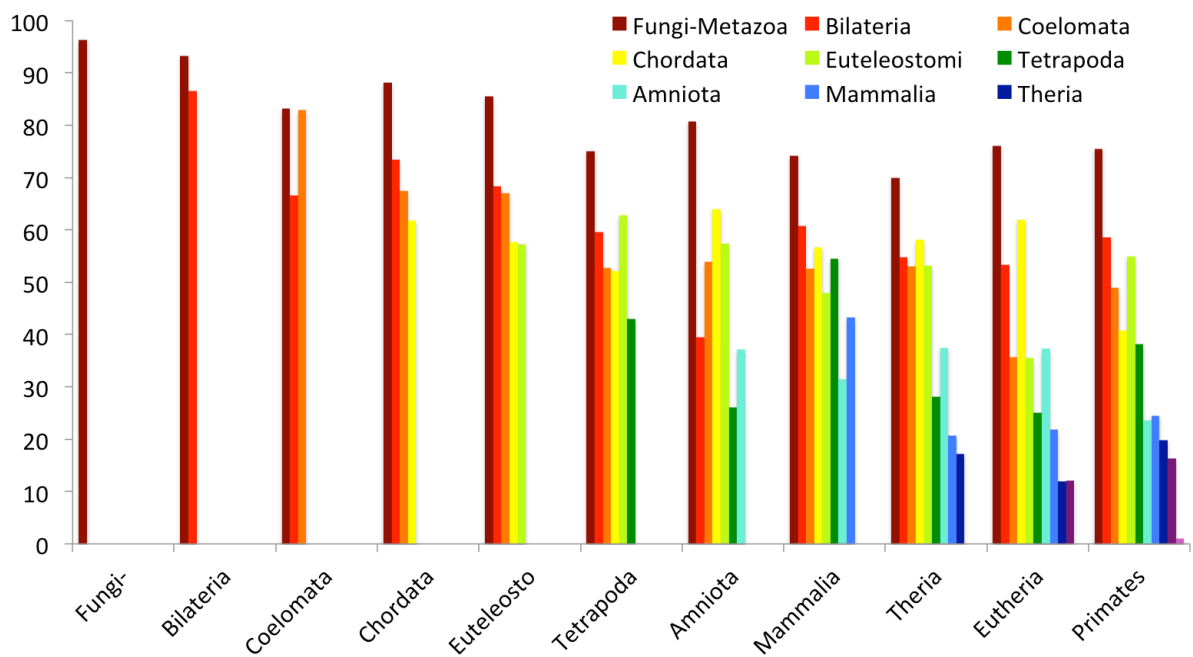

**Figure 4. Gene family age and gene age.**
Here we show the percentage of genes identified in the proteomics studies against ancestral gene age (gene family age) and the age of most recent duplication event (gene age). The y-axis show the percentage of genes detected in proteomics experiments. The x-axis shows the age of the last recorded duplication (oldest duplications on the left, most recent on the right), while the legend shows the gene family age (family ages are also coloured by a rainbow colouring scheme, oldest at the red end of the spectrum, most recent at the blue end). Genes with identical family and duplication age are shown in the furthest right-hand bar of each gene duplication cluster, while those that duplicated from genes with Fungi-Metazoa family age are at the left of each cluster (note that Fungi-Metazoa has no separate gene duplicate bar). Curiously we found that there was significantly more evidence for genes with duplication events that arose from genes with Fungi-Metazoa gene family age than for any other, suggesting that genes that duplicate from the most ancient genes tend to find niches in the cellular processes more easily.

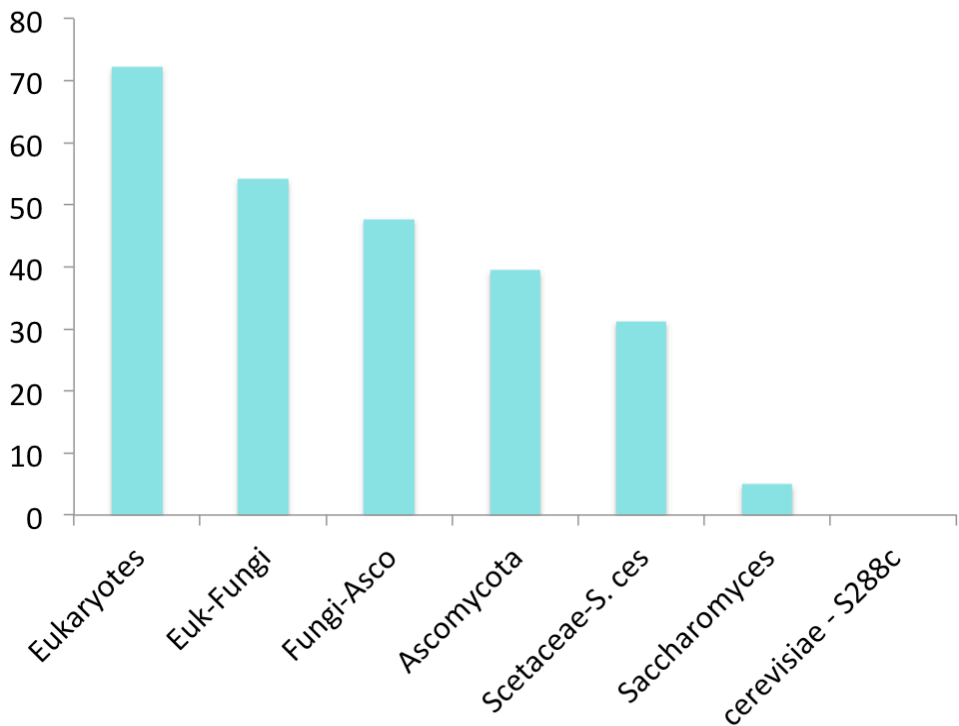

**Figure 5. The relationship between gene family age and peptide detection for yeast.**
Percentage of *Saccharomyces cerevisiae S288c* genes for which peptides detected in proteomics experiments against ancestral gene age, oldest ancestral genes (on the left) are detected much more often in proteomics experiments. PeptideAtlas detects peptides for 73% of yeast genes. There is the same clear relation between gene family age and peptide detection rates in yeast. PeptideAtlas detects no peptides for genes that arose since the *cerevisiae* division. Yeast is a single-celled organism, so here tissue specificity is not the reason why peptides are not detected for certain genes.

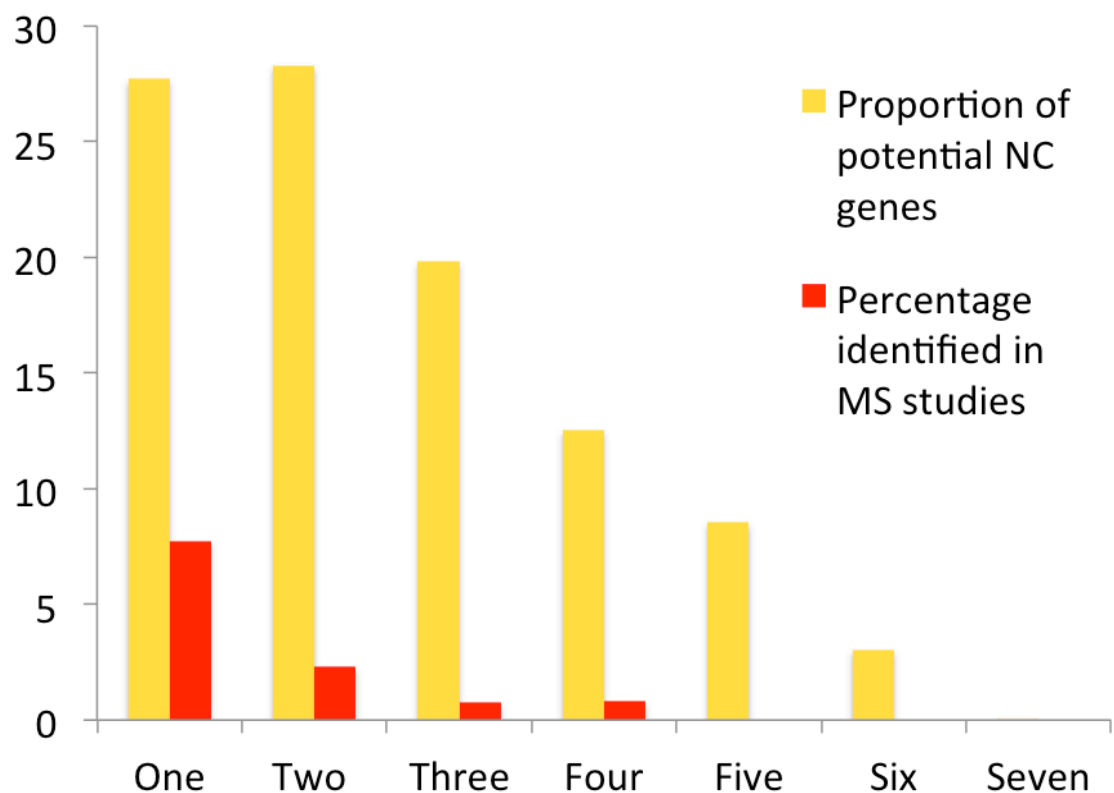

**Figure 6. Number of atypical protein features per gene**
The x-axis shows the number of protein-atypical features per gene in the potential non-coding set, and the y-axis shows (in yellow) the proportion of the 2,001 potential non-coding genes that have the number of atypical features and (in red) the distribution of the 61 genes that we identified in proteomics studies.

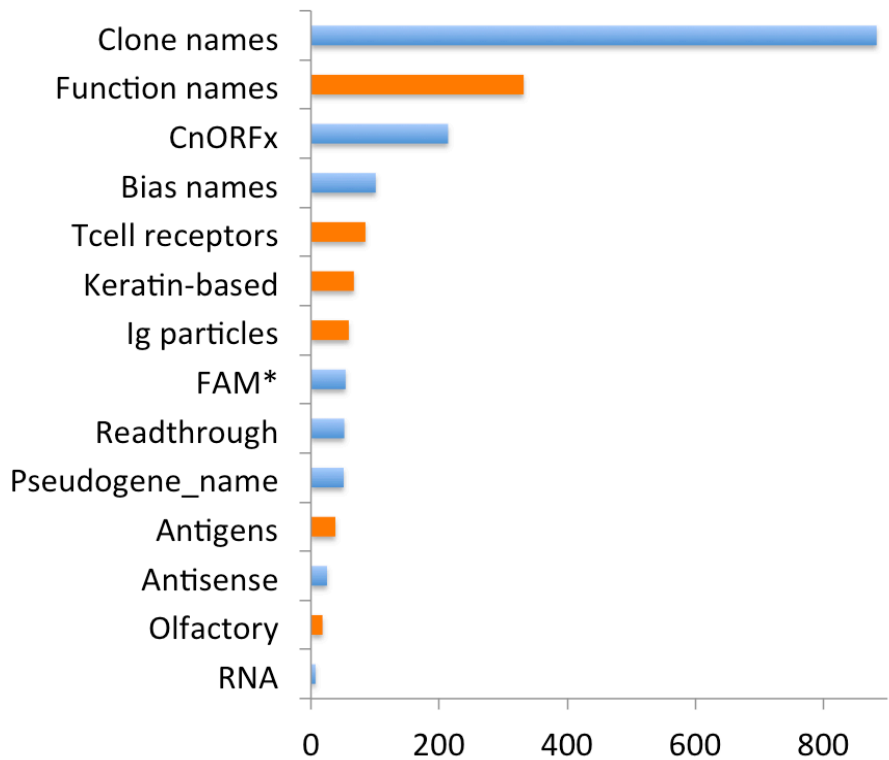

**Figure 7. Gene names in the potential non-coding set**
HGNC/GENCODE naming information for the 2,001 genes in the potential non-coding set. Names were binned according to type. Bars in blue indicate protein names related to position in the genome or amino acid bias; those in orange show gene names that are at least slightly related to potential gene function.

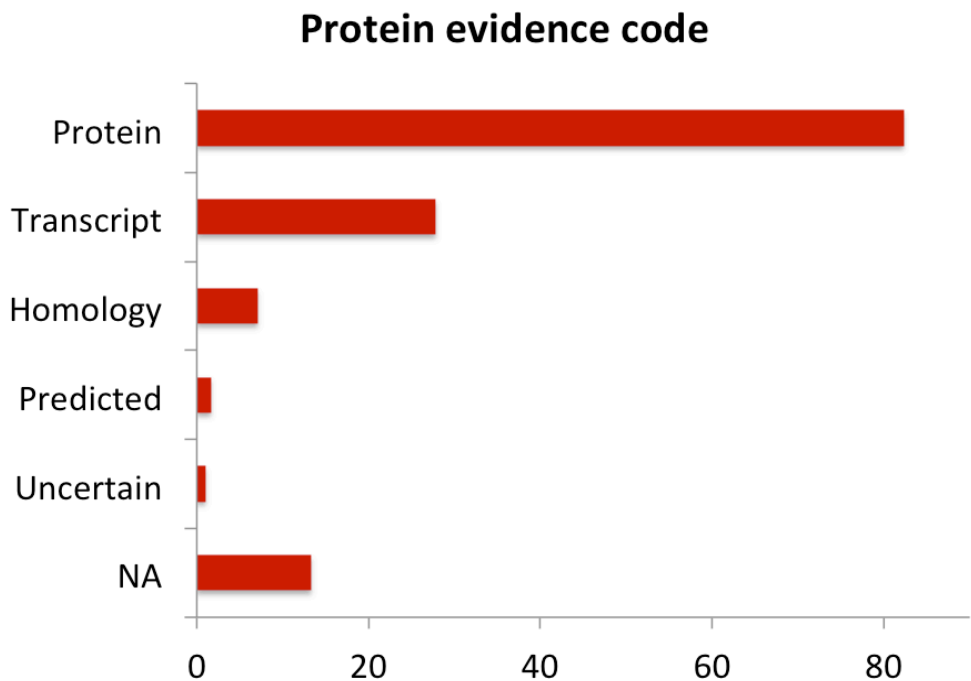

**Figure 8. UniProt protein evidence for GENCODE/Ensembl genes.**

The percentage of genes identified in the proteomics analysis for different levels of UniProt protein evidence. We took the isoform with the best evidence for each gene. "NA" are those genes for which we could not identify an evidence code in UniProt.

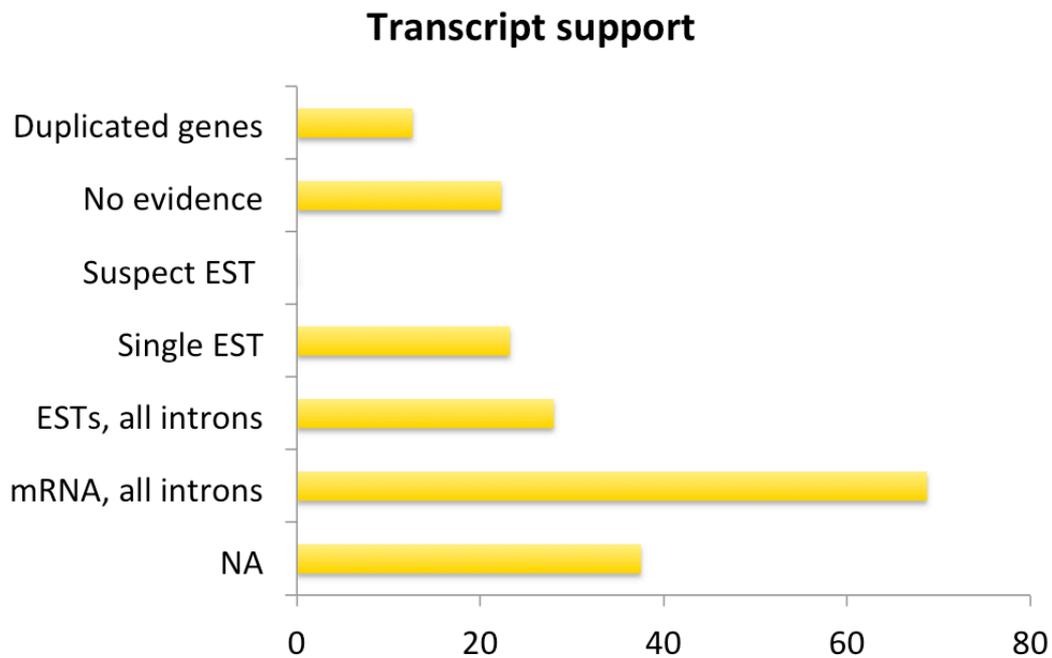

**Figure 9. Transcript support for GENCODE/Ensembl genes.**
The percentage of genes identified in the proteomics analysis for different levels of transcription support. We took the isoform with the best support to represent each gene. "NA" are those genes for which we were not able to find transcription support.

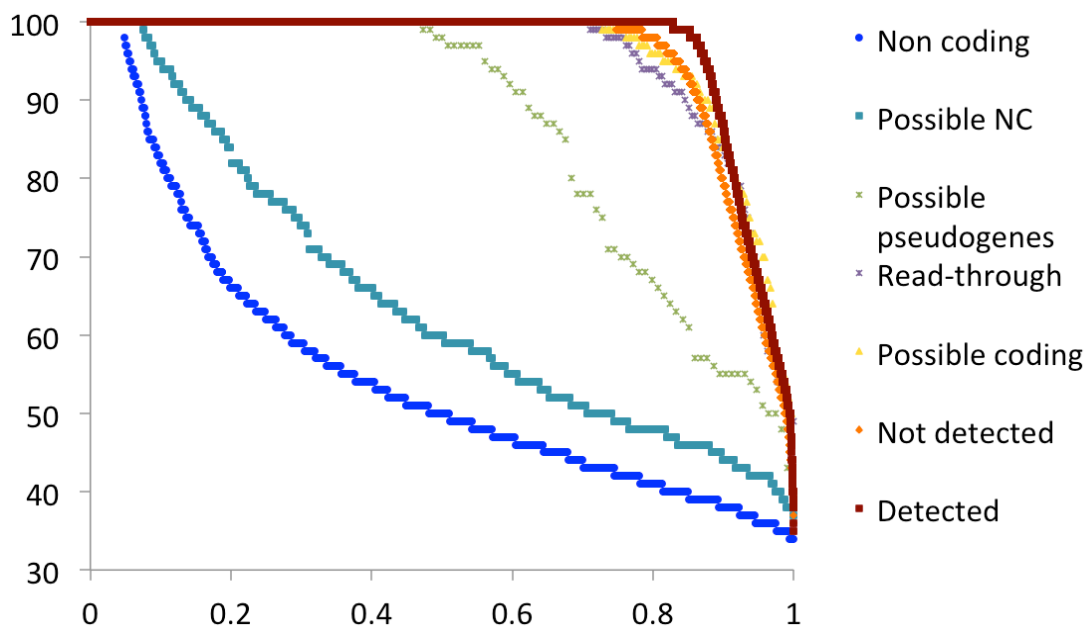

**Figure 10. RFC scores for the classified genes from potential NC set**
The RFC scores were calculated as per the methods section for alignments between human and mouse only. We split the *Potential NC* set genes into 4 groups, those 392 genes that we felt were certain protein coding genes (*Possible coding*), the 343 genes that we felt were possible pseudogenes (*Possible pseudogenes*), the 229 read-through genes and those 968 genes that we felt were likely to be non-coding (*Possible non-coding*). We compared these four sets against three background sets, those protein-coding genes for which we found peptides (*Detected* in dark red), those coding genes for which we did not find peptides and that were not in the potential non coding set (*Not Detected* genes, in orange) and a set of long non-coding genes (*Non-coding* shown in blue). RFC scores are shown on the y-axis, the x-axis in all the figures is the proportion of all the valid pairwise alignments included in the RFC calculations. RFC scores are ordered from highest to lowest.